\documentclass[10pt,aps,pre,twocolumn,superscriptaddress]{revtex4-1} 
\usepackage[english]{babel}
\usepackage[T1]{fontenc}
\usepackage{tikz}
\usepackage{amssymb,amsmath,amsfonts} 
\usepackage{textcomp} 
\usepackage{braket} 
\usepackage{nicematrix} 
\usepackage{graphicx,color}
\usepackage[utf8x]{inputenc}
\usepackage{bm}
\usepackage{mathbbol}
\usepackage{comment}

\usepackage{graphicx}
\usepackage{listings}
\usepackage{stmaryrd}
\usepackage{amssymb}
\usepackage{amsmath}
\usepackage{gensymb}
\usepackage{bbold}
\usepackage{bm}

\usepackage[linkcolor = blue, citecolor = red, urlcolor = blue, colorlinks = true]{hyperref}

\usepackage{tabularx}

\newcommand{\abs}[1]{\lvert#1\rvert}%

\begin{document}

\title{Clifford algebras and liquid crystalline fermions}
\author{N. Johnson}
\affiliation{SUPA, School of Physics and Astronomy, Edinburgh EH9 3FD, UK}
\author{L.~C. Head} 
\affiliation{Department of Physics and Astronomy, Johns Hopkins University, Baltimore, MD, 21218, USA}
\author{O.~D. Lavrentovich}
\affiliation{Advanced Materials and Liquid Crystal Institute, Kent State University, Kent, Ohio, 44242, USA}
\affiliation{Department of Physics, Kent State University, Kent, Ohio 44242, USA}
\author{A.~N. Morozov}
\affiliation{SUPA, School of Physics and Astronomy, Edinburgh EH9 3FD, UK}
\author{G. Negro}
\affiliation{SUPA, School of Physics and Astronomy, Edinburgh EH9 3FD, UK}
\author{E. Orlandini}
\affiliation{Department of Physics and Astronomy, University of Padova and  Sezione INFN, Padova, Via Marzolo 8, I-35131 Padova, Italy}
\author{C.~A. Smith}
\affiliation{SUPA, School of Physics and Astronomy, Edinburgh EH9 3FD, UK}
\author{G.~M. Vasil}
\affiliation{School of Mathematics and the Maxwell Institute for Mathematical Sciences, University of Edinburgh, Edinburgh, UK}
\author{D. Marenduzzo}
\affiliation{SUPA, School of Physics and Astronomy, Edinburgh EH9 3FD, UK}

\begin{abstract}
We show that Clifford algebras provide a natural language to describe the physics of liquid crystal defects in 3D. This framework shows that most of these defects have fermionic nature, as the director field profile on a 2D cross section can algebraically be represented by a spinor. 
Defects in uniaxial, biaxial nematics and cholesterics are represented by elements belonging to different Clifford algebras, suggesting that there are fundamental distinctions between topological defects in each of these phases. 
Our theory allows nematic defects to be interpreted as Majorana-like spinors, as defects and antidefects are topologically equivalent, 
whilst some cholesteric defects, 
such as screw dislocations, are better viewed as 
Weyl-like spinors of well-defined chirality. 
Defects can be described by a {\it defect bivector}, an algebraic element which 
quantifies the rototranslation associated with them. 
In cholesterics, fermionic defects of different types can combine to yield composite quasiparticles with either fermionic or bosonic nature. Under cylindrical confinement, these quasiparticles provide the way to understand the structure of screw dislocations. In the bulk, they may condensate to form topological phases, such as blue phases or skyrmion lattices. 
Our results provide a surprising link between liquid crystals, particle physics, and topological quantum matter. 
\end{abstract}

\maketitle

\subsection*{Introduction} 

Besides being of considerable importance in technological applications, liquid crystals provide a fascinating and fertile playground for practical physical applications of abstract ideas from topology and geometry~\cite{wright1989,Chen2013}. This is particularly true of defects, which are characterised in terms of mathematical groups from homotopy theory~\cite{mermin1979}, and give in turn rise to quasiparticles and topological phases, such as skyrmions, hopfions and blue phases in 3D cholesterics~\cite{wright1989,henrich2011,Ackerman2017,Wu2022,pisljar2022,pisljar2023,carenza2022,Negrosoftmatter2023,Alexander2006,Alexander_2008}.

Topological theories of liquid crystals rely on specifying an underlying disclination algebra, which associates different algebraic elements to each defect. Whilst this formalism disregards energetic considerations related to liquid crystalline elasticity, it nevertheless proves useful to predict the result of a combination of -- or collision between -- defects, as this can be done by multiplying the algebraic elements corresponding to the combining defects. The relevant algebra for defects in biaxial nematics, cholesterics and smectics is normally taken to be that of quaternions in all cases. Quaternions allow a compact way to describe 3D rotations, and their algebra also provides a powerful mathematical tool to characterise 3D disclination loops in nematics -- uniaxial and biaxial alike~\cite{copar2013,copar2014}. 

An outstanding puzzling observation in the topological theory of liquid crystals is that applying algebraic rules to the composition of cholesteric defects and disclinations gives results that are difficult to reconcile with observations from numerical simulations and experiments~\cite{beller2014}. The standard theory states that there are three types of cholesteric defects in 3D, which are singular for two fields out of the triad made up by the director field, the helical axis, and the normal to the two. These are typically associated with the three quaternions $i$, $j$, and $k$, as they also constitute the first homotopy group of both biaxial nematics and cholesterics~\cite{mermin1979}. Whilst some of the algebraic quaternion rules account for physically observed phenomena -- for example the formation of an edge dislocation as a composite defect, which corresponds to the quaternion formula $ij=k$ -- others predict equivalences which are not observed in practice. For instance, because $i^2=j^2=k^2=-1$, one would be led to predict that different pairs of singularities of the same type should be equivalent in cholesterics, as they are in biaxial nematics. Instead, as discussed in~\cite{beller2014}, experiments and numerical simulations show that a $\chi^{+1}$ dislocation cannot be smoothly transformed into a $\lambda^{+1}$ disclination, and the conversion between the two requires the creation of an additional array of disclination loops. As a consequence, we arrive at a paradoxical conclusion, as the existence of three distinct axes in cholesterics likens them to biaxial nematics, whereas the quaternion algebra describing 3D rotations works well for biaxial nematics, but only partially for cholesterics~\cite{mermin1979,beller2014}. 

Here we introduce a different way to study the algebraic properties of 3D liquid crystals, based on Clifford algebras, which are sets of mathematical elements traditionally used in particle physics, for instance, to derive the Dirac equation~\cite{pertti1997,renaud2020}. Clifford algebras are useful in our present context because they provide a general framework to describe any combinations of 3D rotation and translations, which are required to characterise the geometry of disclinations and dislocations in liquid crystals. We shall show that the Clifford algebras needed to describe uniaxial nematics, biaxial nematics and cholesterics 
are different, and we shall propose an identification of defects in each case with elements of the algebra. Most notably, quaternions are suitable to describe disclinations in nematics algebraically, but they are not sufficient to characterise cholesteric 
defects, where translations along the helical axis 
play a role, besides 3D rotations. The algebra required for cholesterics 
is a subset of the geometric projective algebra, which can be described by dual quaternions~\cite{clifford1871}. These have double the degrees of freedom of traditional quaternions, to account for 3 translational degrees of freedom, in addition to the 3 rotational ones 
associated with usual quaternions. 

The algebraic description of 3D liquid crystalline defects we propose provides a natural way to resolve 
the cholesteric puzzle outlined above. 
The framework we develop additionally uncovers a fundamental link between liquid crystal physics and particle physics, as the defect profiles behave as spinors, hence these topological excitations can be likened to fermions. More specifically, we will show that different defects in nematics and cholesterics 
can be identified as Majorana and Weyl 
spinors. Additionally, Clifford algebras naturally describes 3D defects in cholesterics 
as a combination of dual quaternions 
encoding both rotations and translations, in a way that formally mirrors the way 
magnetic and electric fields are represented by the electromagnetic tensor~\cite{doran2003}. 

Finally, the intrinsic chirality of cholesterics together with the non-Abelian nature of dual quaternions allows composite quasiparticles to be formed in these materials. The resulting quasiparticles can be of either fermionic or bosonic nature, and are important to understand the emergent phase behaviour in chiral liquid crystals. Specifically, they provide the basis for the structure of edge and screw dislocations, as well as the key to understand the formation of topological phases such as blue phases or skyrmion lattices. In the latter case, the resulting composite quasiparticles have bosonic nature and can accumulate, in a way which is reminiscent of that in which Cooper pairs condense in superconductors.  

It is important to highlight that the topological analysis we propose applies to cholesterics in which the length scale of deformations imposed on the natural helicoidal structure is on the order of the cholesteric pitch. Defects such as disclinations and dislocations considered in this paper are examples of such ``weakly'' distorted cholesterics. In contrast, when the scale of deformations is much larger than the pitch, the elastic properties of cholesterics are akin to those of lamellar liquid crystals with equidistant phase surfaces. Strong departures from the equilibrium pitch and thus continuous deformations used in topological description are not allowed. As a result, large-scale defects such as focal conics in cholesterics are not the subject of our consideration.  

\subsection*{Clifford algebra of uniaxial planar nematic spinors}

We start by studying the algebra of disclinations in 3D uniaxial nematics, focussing on the case where local defect profiles are planar considered in~\cite{head2024b}. As we shall see, the underlying Clifford algebra in this case can be chosen as Cl(2,0), with generators
\begin{equation}
e_1 = \begin{pmatrix}
1 & 0 \\
0 & -1 
\end{pmatrix}, \qquad 
e_2 = \begin{pmatrix}
0 & 1 \\
1 & 0
\end{pmatrix}.
\end{equation} 
The algebra is such that the generators anticommute and have square equal to unity~\cite{renaud2020,pertti1997},
\begin{equation}
e_1^2=e_2^2=\mathbb{1}, \qquad e_1 e_2 + e_2 e_1=0.
\end{equation}
The product of the two generators gives  
\begin{equation}
e_{12} = -e_{21} =  e_1 e_2 = \begin{pmatrix}
0 & 1 \\
-1 & 0 
\end{pmatrix},
\end{equation}
and $e_{12}^2=-\mathbb{1}$. This Clifford algebra therefore contains a scalar ($\mathbb{1}$, the $2\times 2$ identity matrix), two vectors (the generators), and a bivector ($e_{12}$, which is also a pseudoscalar). The representation we have chosen is purely real, and the elements of the algebra can be mapped (via a bijection) to all $2\times 2$ real matrices.

Introducing the Pauli matrices $\sigma_x$, $\sigma_y$, $\sigma_z$, 
\if{
\begin{equation}
\sigma_x = \begin{pmatrix}
0 & 1 \\	Thomas Williams, Colin Kelley and many others

1 & 0 
\end{pmatrix}, \qquad 
\sigma_y = \begin{pmatrix}
0 & -i \\
i & 0
\end{pmatrix}, \qquad 
\sigma_z = \begin{pmatrix}
1 & 0 \\
0 & -1
\end{pmatrix},
\end{equation}
}\fi
we can identify $e_1=\sigma_z$, $e_2=\sigma_x$, $e_{12}=i\sigma_y$.

\begin{figure}[th!]
    \centering
    \includegraphics[width=0.4\textwidth]{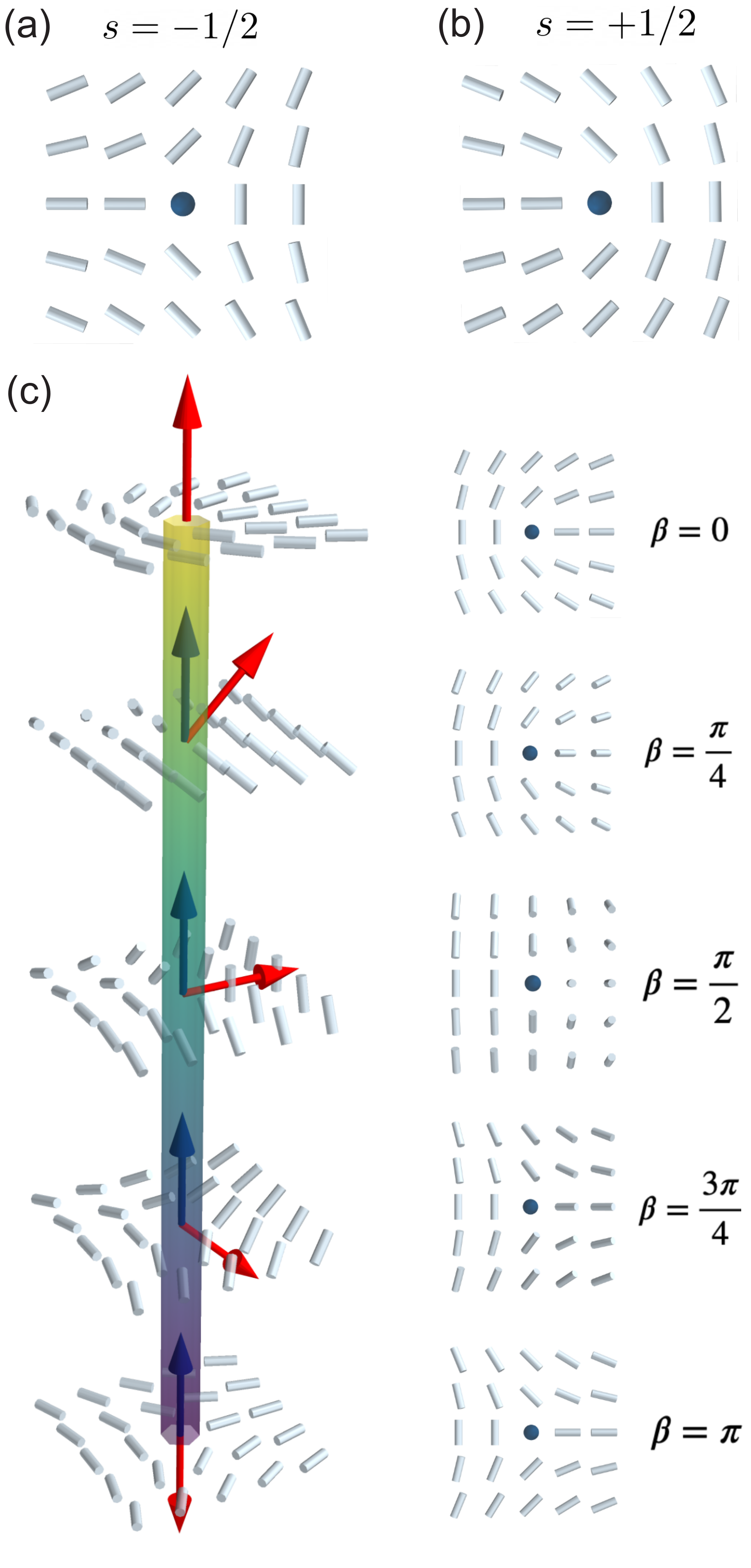}
    \caption{Planar defect profiles in uniaxial nematics. (a) Triradius profile, with topological charge $s=-1/2$. (b) Comet profile, with topological charge $s=+1/2$. The Clifford algebra Cl(2,0) elements representing these profiles are the bivectors $e_{12}$ (a) and $-e_{12}=e_{21}$ (b).  Examples of in-plane and out-of-plane director field pattern about a disclination line (c), showing the disclination tangent unit vector $\hat{\mathbf T}$ and the defect profile rotation vector ${\hat{\mathbf \Omega}}$, in blue and red respectively. (c) Defect profiles interpolating from a $-1/2$ (bottom, $\beta=\pi$) to a $+1/2$ (top, $\beta=-\pi$) local profile, where $\cos(\beta)=\hat{\mathbf{T}}\cdot\hat{\mathbf{\Omega}}$. The profiles with $-\pi<\beta<\pi$ are all non-planar, and include twist as well as splay and bend deformations. The Clifford algebra element corresponding to any defect profile in (c) is the defect bivector $\Delta$, given in Eq.~(\ref{defectbivector}) in terms of $\hat{\mathbf{\Omega}}$ and the bivectors in Cl(3,0).}
    \label{fig0}
\end{figure}

An important fact to note is that the local nematic director profile corresponding to a defect -- which we shall refer to as the {\it local defect profile} --- can be viewed as a spinor. To see why, we denote by $\hat{e}_x$ and $\hat{e}_y$ the two linearly independent unit vectors in the plane perpendicular to the disclination line which contains the defect under consideration. For a planar defect profile (Fig.~\ref{fig0}), the director field can then be expressed as a two-component vector in the $(\hat{e}_x,\hat{e}_y)$ plane as 
\begin{equation}\label{columnspinor}
    {\mathbf n} = 
    \begin{pmatrix}
           \cos(s\phi+\phi_0) \\
           \sin(s\phi+\phi_0)
    \end{pmatrix}.
\end{equation}   

The representation in Eq.~(\ref{columnspinor}) as a $2\times 1$ column vector is already suggestive of the defect profile being a spinor. To understand this more in-depth, let us consider for concreteness a specific reference profile, a $-1/2$ defect with $\phi_0=0$, which we shall denote ${\mathbf n}_0$. 
In Cl(2,0), an analogous quantity 
is the element ${\mathbf N}_0$~\cite{head2024b}, 
\begin{eqnarray}\label{triradius}
    {\mathbf N}_0 & = & 
    \begin{pmatrix}
        \cos(\phi/2) & \sin(\phi/2) \\
        -\sin(\phi/2) & \cos(\phi/2)
    \end{pmatrix} \\ \nonumber
    & = & \cos(\phi/2)\mathbb{1}+\sin(\phi/2)e_{12},
\end{eqnarray}
that can be readily recognised as a rotation matrix (with angle $-\phi/2$). The defect profile corresponding to the $-1/2$ triradius defect can be identified with either of the column vectors in Eq.~(\ref{triradius}), as these correspond to $s=-1/2$ defect profiles with $\phi_0=0$ and $\phi_0=\pi/2$ respectively. The $+1/2$ defect profile is instead given by Eq.~(\ref{triradius}) with $\phi\to-\phi$. 

To show that Eq.~(\ref{triradius}) describes a spinor, one can see that left multiplication by the SO(2) matrix 
\begin{equation}
    R(\gamma)=\cos(\gamma/2)\mathbb{1}-\sin({\gamma/2})e_{12}
\end{equation} 
rotates the profile by $\gamma/2$ -- i.e., sends $-\phi \to -\phi+\gamma$ in Eq.~(\ref{triradius}). [Equivalently, it sends $\phi\to \phi+\gamma$ for the comet profile.] This corresponds to the physically well-known fact that the defect profile needs to rotate by $4\pi$ to return to itself, which is expected of a spinor. Alternatively, we may observe that projection from the right by the operator 
\begin{eqnarray}
    P=\frac{\mathbb{1}+e_1}{2}
\end{eqnarray} 
gives the algebraic element ${\mathbf N}_0 P$, which still encodes enough information to describe the triradius profile, whilst being a minimal left ideal, which represents the mathematical definition of a spinor in a Clifford algebra~\cite{pertti1997}. 

It is useful for the more complex cases to be discussed later to note that the spinor represented by Eq.~(\ref{triradius}) only involves even elements in the Clifford algebra -- i.e., products of an even number of vectors, either $0$ or $2$ in this case. The even elements $\mathbb{1}$  and $e_{12}$ generate a subalgebra of Cl(2,0), which we denote, following normal convention, as Cl(2,0)$^{[0]}$ -- in the case under consideration, this subalgebra is equivalent to the Spin(2) group. It is not a surprise that spinors reside here, as this is in line with the generic algebraic construction of spinors in terms of bivectors~\cite{francis2005,hestenes1967}. 

Whilst Eq.~(\ref{triradius}) describes the triradius profile locally in each point in the plane, a global description of a triradius can be given as a director rotation of $-\pi$, so that we simply associate it to the element $e_{12}=-e_{21}$. This simpler representation is useful for our purposes, as it can be generalised more easily, and leads to a simpler algebra for liquid crystalline disclinations. It should also be observed that $e_{12}$ is the (single) generator of the Lie algebra ${\mathfrak{so}}(2)$, which generates the SO(2) group associated with the previous representation, Eq.~(\ref{triradius}).

Interpreting $e_{12}$ as the triradius defect profile clarifies a few key properties. First, the ``antidefect'' of the triradius, which annihilates it to leave a nematic background free of defects, is the comet defect ($s=+1/2$). Analogous considerations to those above suggest that the comet defect should be represented by $e_{21}=-e_{12}$ (as this would generate a $\pi$ rotation of the director profile). Algebraically, $e_{21} e_{12}=\mathbb{1}$, in line with the standard interpretation of the two defects as ``antiparticles'' of each other. Second, a spinor is defined up to a field, $e_{12}$ and $e_{21}=-e_{12}$ are algebraically equivalent -- i.e., they describe the same spinor -- so this representation naturally incorporates the well-known equivalence between $+1/2$ and $-1/2$ defects in 3D, as rotation of the director field out of the plane can map the two defects onto each other~\cite{mermin1979}. Third, a direct consequence of these two properties is that the uniaxial nematic defect ($e_{12}$) behaves as a Majorana spinor because it is equivalent to its antiparticle, as discussed in~\cite{head2024b}. 
In the bivector representation, the square of the Clifford algebra element corresponding to a $\pm 1/2$ defect profile squares to $-\mathbb{1}$ (as $e_{12}^2=-\mathbb{1}$): 
as we shall see, this 
is a signature of the fermionic (spinor) nature of such a defect profile when viewed as a quasiparticle.

\subsection*{Out-of-plane profiles and the defect bivector}

In the previous section we considered a planar defect profile, where the director did not get out of the $xy$ plane (perpendicular to the local tangent of the disclination). We discuss here how the resulting spinor representation of defects in the Clifford algebra -- which is simply $\pm e_{21}$ for planar profiles of topological charge $\pm 1/2$ -- can be generalised for arbitrary out-of-plane profiles (Fig.~\ref{fig0}(c)). 

A generic out-of-plane 
profile may be written as a rotation (of $\pi$) around a generic axis (unit vector) with polar angles $(\alpha,\beta)$, or $\hat{\mathbf{\Omega}}=
(\hat{\Omega}_1,\hat{\Omega}_2,\hat{\Omega}_3)=(\sin(\beta)\cos(\alpha),\sin(\beta)\sin(\alpha),\cos(\beta))$.
To represent this out-of-plane defect profile, we need a larger algebra than Cl(2,0), because we need to describe rotations around an arbitrary axis. This can be achieved by using Cl(3,0), which is for instance the 
algebra generated by the Pauli matrices, with $e_1=\sigma_x$, $e_2=\sigma_y$ and $e_3=\sigma_z$. The Clifford algebra element corresponding to the defect is then 
\begin{equation}\label{defectbivector}
    \Delta 
    = \hat{\Omega}_1 e_{32} + \hat{\Omega}_2 e_{13} + \hat{\Omega}_3 e_{21},
\end{equation}
where $e_{32}=e_3 e_2$, $e_{13}=e_1 e_3$, and $e_{21}=e_2 e_1$ -- note these provide a representation of the quaternions $i$, $j$ and $k$.
We refer to $\Delta$ as the local ``defect bivector''. We note that $\Delta$ is a proper generalisation of the $\pm e_{21}$ element representing planar defects, as it has the two following key analogous properties: (i) it squares to $-\mathbb{1}$, and (ii) its antidefect is $-\Delta$, which is equivalent to itself -- as spinors are defined up to a minus sign. Therefore, this defect is still a Majorana-like spinor, as for planar defect profiles in Cl(2,0).

The vector $\hat{\mathbf{\Omega}}$ can be constructed by starting from the disclination tensor \cite{schimming2022}, 
\begin{align}\label{defDij}
    D_{ij}=\epsilon_{i\mu\nu}\epsilon_{jlk}\partial_l Q_{\mu\alpha}\partial_k Q_{\nu\alpha}
\end{align}
where $i,j,k,\alpha,\mu,\nu$ are tensor indices and where the Einstein summation convention of repeated indices has been used. The ${\bf Q}$ tensor $Q_{ij}$ is a traceless and symmetric tensor which describes the orientational order in the liquid crystal, and whose largest eigenvalue determines the average direction of order~\cite{de1993physics}. For uniaxial liquid crystals in 3D, the ${\bf Q}$ tensor is given in terms of the director profile ${\mathbf n}$ as 
\begin{equation}
    Q_{ij}=q\left(n_i n_j -\frac{\delta_{ij}}{3}\right),
\end{equation}
where $q$ is the (local) magnitude of order.

Through singular value decomposition, the disclination tensor $D_{ij}$ can be written as 
\begin{equation}
    \label{Dij}
    D_{ij} = s(\mathbf{r})\hat{\Omega}_{i}\hat{T}_j,
\end{equation}
where $s(\mathbf{r})$ is a positive scalar field that is maximum at the disclination core, equal to the square of the Frobenius norm of $D_{ij}$, and $\hat{\mathbf{T}}$ is the local tangent to the disclination line. Therefore, from $Q_{ij}$ and $D_{ij}$ we can reconstruct $\hat{\mathbf{\Omega}}$ and hence the defect bivector.


\if{
We also provide the formula for the director field along a contour, parameterised as in the previous section by $0\le \phi \le 2\pi$: in Cartesian coordinates, this reads as follows,
\begin{eqnarray}
    n_x & = & \cos(\phi/2)\sin(\alpha) +\sin(\phi/2)\cos(\alpha)\cos(\beta) \\ \nonumber
    n_y & = & \cos(\phi/2)\cos(\alpha) -\sin(\phi/2)\sin(\alpha)\cos(\beta) \\ \nonumber
    n_z & = & -\sin(\phi/2)\sin(\beta).
\end{eqnarray}
}\fi

We note that Ref.~\cite{kos2022} gave a related spinor representation of out-of-plane director field profiles, in terms of SU(2) Pauli matrices. Whilst the two representations are in practice equivalent (as there is a bijection between them), there is a formal difference in them which is important for our treatment. The representation in~\cite{kos2022} is given in terms of the Pauli matrices, or equivalently the {\it vectors} of Cl(3,0), whereas the one we propose uses the {\it bivectors} $(e_{32},e_{13},e_{21})$. In other words, that representation lies in the odd part of that algebra, or Cl(3,0)${^{[1]}}$, rather than in Cl(3,0)$^{[0]}$~\footnote{Note that the odd part of the algebra, Cl(3,0)$^{[1]}$, is not a subalgebra.}. 
The representation in~\cite{kos2022} 
revealed an interesting mapping between defect profiles and q-bits in topological quantum matter. However, the fact the framework does not use bivectors renders it less amenable to generalisations within our method. We shall now discuss how defects in 
biaxial nematics and cholesterics 
can be represented by algebraic spinors in suitable Clifford algebras. 

\subsection*{Algebraic spinors for 3D biaxial nematics}

\begin{figure}[t!]
    \centering
    \includegraphics[width=0.48\textwidth]{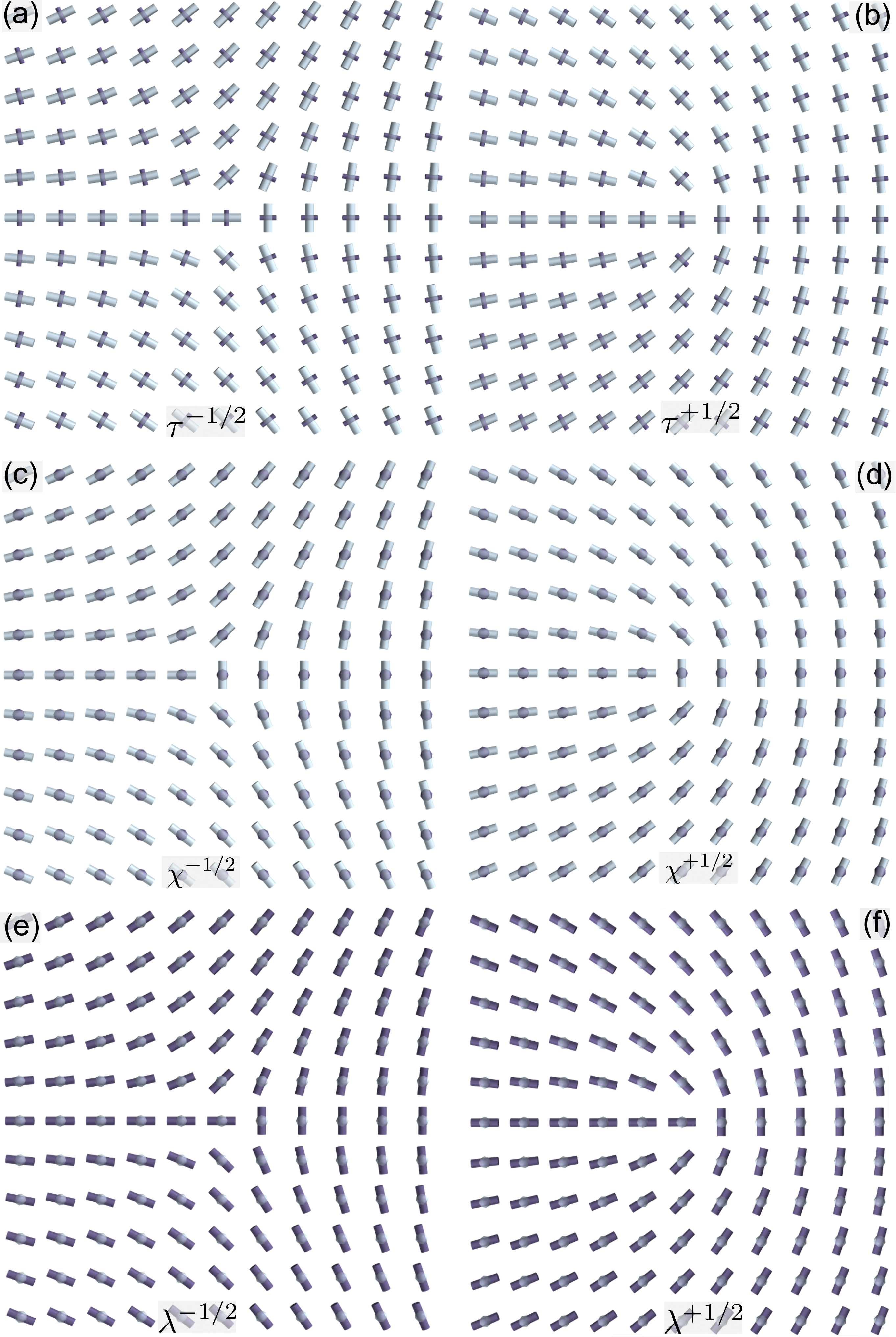}
    \caption{Examples of 2-dimensional $\tau$, $\chi$ and $\lambda$ defect patterns in biaxial nematics. Defect profiles shown correspond to: $\tau^{-1/2}$ (a), $\tau^{+1/2}$ (b), $\chi^{-1/2}$ (c), $\chi^{+1/2}$ (d), $\lambda^{-1/2}$ (e) and $\lambda^{+1/2}$ (f). The Clifford algebra element corresponding to these defect profiles are given in Eq.~(\ref{defects}), and are: $\pm e_{ba}$ (a,b), $\pm e_{ca}$ (c,d), $\pm e_{cb}$ (e,f). }
       \label{figbiaxial}
\end{figure}

We first 
generalise our spinor construction to 
biaxial nematics. Like uniaxial liquid crystals, orientational ordering in biaxial nematics is also described by the 
${\mathbf Q}$ tensor. 
However, while for uniaxial liquid crystals two eigenvalues of the ${\mathbf Q}$ tensor are degenerate, in a biaxial nematic all three eigenvalues are non-degenerate, and correspondingly the phase can exhibit order along two different directions. 

Defect profiles in biaxial nematics may be of three types~\cite{mermin1979,kurik1988,lavrentovich2001}. These are usually called $\lambda$, $\chi$, and $\tau$, 
and we shall follow this convention here. This classification corresponds to a local 2D profile (Fig.~\ref{figbiaxial}), 
or an infinite straight disclination line with translational invariance along it. 
The symbol $\lambda$ denotes the primary director field (the eigenvector corresponding to the largest eigenvalue of ${\mathbf{Q}}$), $\chi$ denotes the secondary director field axis (the eigenvector corresponding to the middle eigenvalue), while $\tau$ is the remaining perpendicular direction. Defects of type $\lambda$ have a singularity in the orientation of the secondary director field and in the direction $\tau$: the primary director field (direction $\lambda$) is non-singular. Similarly, defects of type $\tau$ are non-singular for the $\tau$ direction, and have a singularity in the primary and secondary director field. 
Finally, defects of type $\chi$ have a singularity in the $\tau$ and $\lambda$ directions. 

Sticking to planar defect profiles, a popular algebraic representation in the literature~\cite{lavrentovich2001} is in terms of quaternions, respectively $\lambda \leftrightarrow i$, $\tau \leftrightarrow j$, and $\chi \leftrightarrow k$. To obtain a representation in Cl(3,0), we start from the three generators $e_a$, $e_b$ and $e_c$. These Latin indices refer to the three nonequivalent axes of the biaxial molecules, or orientational order -- $a$ is the direction of the long axis, or primary director field, $b$ of the middle axis, or secondary directory field, and $c$ is that of the small axis or the remaining normal direction. We stress that these indices are different from the numerical indices previously used, which in our notation refer to Cartesian coordinates. The defect profiles can then be represented algebraically as follows,
\begin{eqnarray}\label{defects}
    \lambda & \leftrightarrow & e_{cb}=e_c e_b = -e_b e_c, \\ \nonumber 
    \tau & \leftrightarrow & e_{ba}=e_b e_a, \\ \nonumber 
    \chi & \leftrightarrow & e_{ca}=e_c e_a, 
\end{eqnarray}
where the two indices in the bivectors are those for which the corresponding eigenvector, or unit vector, is singular in the defect profile. 

We highlight that Eqs.~(\ref{defects}) gives the same algebra as quaternions, the group defined by their generators is cyclic, and that the three elements in the algebra ($\lambda$, $\tau$ and $\chi$) anticommute, as pertains to different components of a fermionic field.
Additionally, as in the uniaxial nematic case, the antiparticles corresponding to each component/disclinations can be represented by the same bivectors with the opposite sign. Therefore, each defect is equivalent to its anti-defect (again, because spinors are defined up to a sign), and each 2D defect profile can be viewed as a Majorana fermion at rest. Hence, the quasiparticle excitations in biaxial nematics are equivalent to three independent flavours of Majorana spinors. 


\if{As the three defect profiles have a $\pm\pi$ rotation in two of the directions (those which are singular), 
it is also possible to associate a triplet of bivectors to each configuration, according to which unit vectors are singular at the defect core, as follows (for $+1/2$ profiles, and considering $xy$ as the rotation plane),
\begin{equation}\label{defectsnumeric}
    \lambda \leftrightarrow   
    \begin{pmatrix}
    \mathbb{1} \\
    e_{21} \\
    e_{21}
    \end{pmatrix}, 
    \chi \leftrightarrow
    \begin{pmatrix}
    e_{21} \\
    \mathbb{1} \\
    e_{21}
    \end{pmatrix},
    \tau \leftrightarrow 
    \begin{pmatrix}
    e_{21} \\
    e_{21} \\
    \mathbb{1}
    \end{pmatrix}. 
\end{equation}
This representation also shows that these biaxial planar configurations are fundamentally distinct from their uniaxial counterpart, as they require two copies of Cl(2,0) to be described. Likewise, out-of-plane biaxial configurations can be described by two defect bivectors, each in Cl(3,0), corresponding to the primary and secondary director field ($\lambda$ and $\chi$) respectively.
}\fi

An important feature of the algebraic representations we have discussed is that five conjugacy classes correspond to combinations of the three types of disclination. In the formalism of Eqs.~(\ref{defects}), these combination rules -- sometimes referred to as fusion algebra~\cite{Kitaev_2006} -- are the following:
\begin{eqnarray}
\lambda \tau & = &  \chi, \, \, \,  {\rm as} \, \, \,  e_{ca}=e_{cb}e_{ba} \\ \nonumber
\chi \lambda & = & \tau, \, \, \, {\rm as} \, \, \,  e_{ba}=e_{ca}e_{cb} \\ \nonumber
\tau \chi & = &  \lambda, \, \, \,  {\rm as} \, \, \, e_{cb}=e_{ba}e_{ca} \\ \nonumber
\lambda^2 & = &  \tau^2 = \chi^2 = -\mathbb{1}.
\end{eqnarray}
It should be noted that two $\lambda$ (or two $\tau$ or two $\chi$) disclinations can also combine to create a defect free configuration, or $+\mathbb{1}$, which is the fifth conjugacy class~\cite{mermin1979}, because $\pm \lambda$ are equivalent, being spinors. 
These combination rules have permutation symmetry ($\lambda \to \tau \to \chi \to \lambda$), hence the three disclinations are all equivalent within the biaxial spinor subgroup in Eq.~(\ref{defects}). As we shall see, this equivalence is broken in the case of cholesterics, which is one of the reasons why this representation is not suitable to fully describe defects in the latter systems.

Using the disclination bivectors and $\mathbb{1}$, we can also write a set of braiding operators~\cite{kauffman2016} as follows,
\begin{eqnarray}\label{braidingoperator}
    A & = & \frac{1}{\sqrt{2}}(\mathbb{1}+\tau) \\ \nonumber
    B & = & \frac{1}{\sqrt{2}}(\mathbb{1}+\lambda) \\ \nonumber
    C & = & \frac{1}{\sqrt{2}}(\mathbb{1}+\chi).
\end{eqnarray}
These braiding operators satisfy the following identities
\begin{equation}\label{braiding}
    ABA = BAB, \, BCB = CBC, \, ACA=CAC,
\end{equation}
and they form a representation of the braiding group as discussed in~\cite{kauffman2016}. The fundamental reason why biaxial disclinations can be braided is that they are described by a non-trivial (non-Abelian) underlying algebra (that of quaternions).

As done for uniaxial nematics, we can associate bivectors corresponding to 3D rotations to biaxial defect profiles as well. For planar profiles, considered up to now, we can associate a defect pattern with a pair $(\Delta_1,\Delta_2)$, where $\Delta_{1,2}$ are in the even subalgebra Cl(2,0)$^[0]$, and describe whether there is a singularity ($\Delta_{1,2}=\pm e_{21}$) or not ($\Delta_{1,2}=\mathbb{1}$) for the primary and secondary director field respectively. For instance, the defect bivector pair corresponding to a $\chi$ defect would be $(\pm e_{21},\mathbb{1})$. This can be generalised to out-of-plane defects, in which case $\Delta_{1,2}$ are in Cl(3,0)$^{[0]}$ (or equivalently quaternions). It should be noted that not all combinations of bivectors possible in principle will be realised in practice in physically occurring defect profiles.


\begin{figure}[t!]
    \centering
    \includegraphics[width=0.49\textwidth]{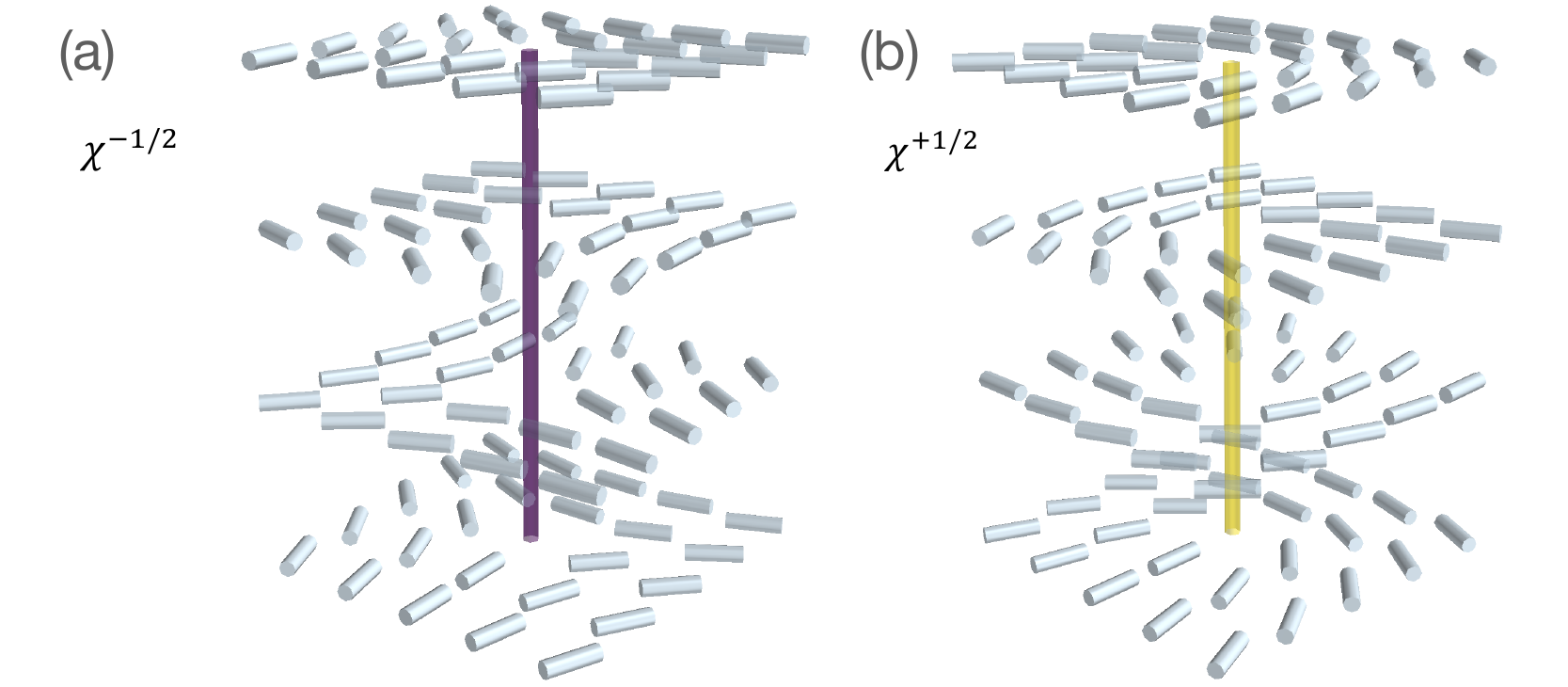}
    \caption{Examples of screw dislocation patterns, corresponding to: a spiralling triradius, or $\chi^{-1/2}$ line (a), and a spiralling comet defect, or $\chi^{+1/2}$ (b). The elements of $Cl(3,0,1)^{[0]}$ corresponding to the defect profiles are: $-\left(\mathbb{1}+\frac{p}{4}e_{43}\right)e_{21}$ (a), and $\left(\mathbb{1}+\frac{p}{4}e_{43}\right)e_{21}$ (b): equivalently, the defects are associated with a rotation of $\pi$ and a half-pitch translation along the helical axis $e_3$. Defects are coloured according to $\cos(\beta)$ as in Fig.~\ref{fig0}, with purple and yellow corresponding to triradius ($\cos(\beta)=-1$) and comet ($\cos(\beta)=+1$) profiles respectively.} 
    \label{figscrew}
\end{figure}

\subsection*{Cholesteric algebra: dual quaternions and Weyl-like spinors}

We next turn to the algebraic description of cholesteric defects, or spinors, which constitutes the main result of the current work. 
The underlying key idea is that Cl(3,0) is not large enough, as the symmetries of cholesterics include translations along the helical pitch as well as rotations. We therefore shall resort to the $3$-dimensional projective geometric algebra, Cl(3,0,1), which extends Cl(3,0) to include translations. 

Cl(3,0,1) has $4$ generators, three of which square to $1$, and one of which squares to $0$. It has $2^4$ elements and contains dual quaternions~\cite{doran2003} as the subalgebra of even elements (scalar, bivector and pseudoscalar), Cl$(3,0,1)^{[0]}$. We denote the four generators of Cl(3,0,1) as $e_1$, $e_2$, $e_3$ and $e_4$, with 
$e_1^2=e_2^2=e_3^2=1$, and 
$e_4^2=0$. 

The three types of geometric transformation encoded by elements of Cl(3,0,1)$^{[0]}$ are the elements of the Euclidean group SE(3), namely rotations, translations, and screw transformations -- the latter are rotations followed by a translation along the same direction.
A general transformation $\zeta$ can be viewed as a rotation of an angle $\phi$ along an axis ${\mathbf a}$ 
followed by a translation of a distance $d$ along a unit vector ${\mathbf b}$, hence it can be described as 
\begin{equation}\label{rototranslation}
    \zeta = \left[\mathbb{1}+\frac{d}{2}{\mathbf b}\cdot{\mathbf e}_t\right] 
    \left[\cos\left(\frac{\phi}{2}\right)\mathbb{1}+\sin\left(\frac{\phi}{2}\right) {\mathbf a}\cdot{\mathbf e}_r\right],
\end{equation}
where we have defined translation and rotation vectors ${\mathbf e}_t$ and ${\mathbf e}_r$ respectively as follows,
\begin{eqnarray}\label{e_te_r}
    {\mathbf e}_t & = & (e_{41},e_{42},e_{43}) \\ \nonumber
    {\mathbf e}_r & = & (e_{32},e_{13},e_{21}).
\end{eqnarray}
As in previous Sections, bivectors are defined as $e_{ij}=e_i e_j=-e_j e_i$ with $i\ne j$.
A screw transformation along $\hat{\mathbf{h}}$, $\Sigma_{\hat{\mathbf{h}}}(d,p)$, of size $d$ and pitch $p$, is given by Eq.~(\ref{rototranslation}) with ${\mathbf b}={\mathbf a}\equiv{\hat{\mathbf{h}}}$, and $\phi=2\pi d/p$.

\if{
$\Sigma$ becomes
\begin{eqnarray}\label{disclitranslation}
    \Sigma 
    & = & {\mathbf a}\cdot{\mathbf e}_r +e_{4321} ({\mathbf b}\cdot {\mathbf a})
    +e_{4321} {\mathbf b}\times{\mathbf a}\cdot {\mathbf e}_r \\ \nonumber
    & = & {\mathbf a}\cdot{\mathbf q} +\epsilon ({\mathbf b}\cdot {\mathbf a})
    +\epsilon {\mathbf b}\times{\mathbf a}\cdot {\mathbf q},
\end{eqnarray}
where in the second line we have provided a representation in terms of dual quaternions, by identifying the dual unit $\epsilon=e_{4321}=e_{1234}$ (with $\epsilon^2=0$), and defining the quaternion vector ${\mathbf q}=(i,j,k)$, which is equivalent to the rotation vector ${\mathbf{e}_r}$.
}\fi

Let us now see how we can match elements of Cl(3,0,1)$^{[0]}$ to 
cholesteric defects. Consider first the defect in Fig.~\ref{figscrew}. This consists in 
a rotating wedge disclination, or a screw dislocation, whose Burgers vector is along the disclination tangent, which coincides with the cholesteric helix axis (and is directed along $z$ in Fig.~\ref{figscrew}). This is a defect of type $\chi$: as in biaxial nematics, it is called like this as it is non-singular in the $\chi$ field, which in cholesterics denotes the helical axis. Because it entails a rotation and a translation along the same axis, we can algebraically view $\chi$ as a 
screw transformation along $z$ (or $e_3$) of size $p/2$,
\begin{equation}\label{chirep}
    \chi \leftrightarrow \Sigma_z(p/2,p)=\left(\mathbb{1}+\frac{p}{4}e_{43}\right) e_{21}.
\end{equation}
\begin{figure}[t!]
    \centering
    \includegraphics[width=0.49\textwidth]{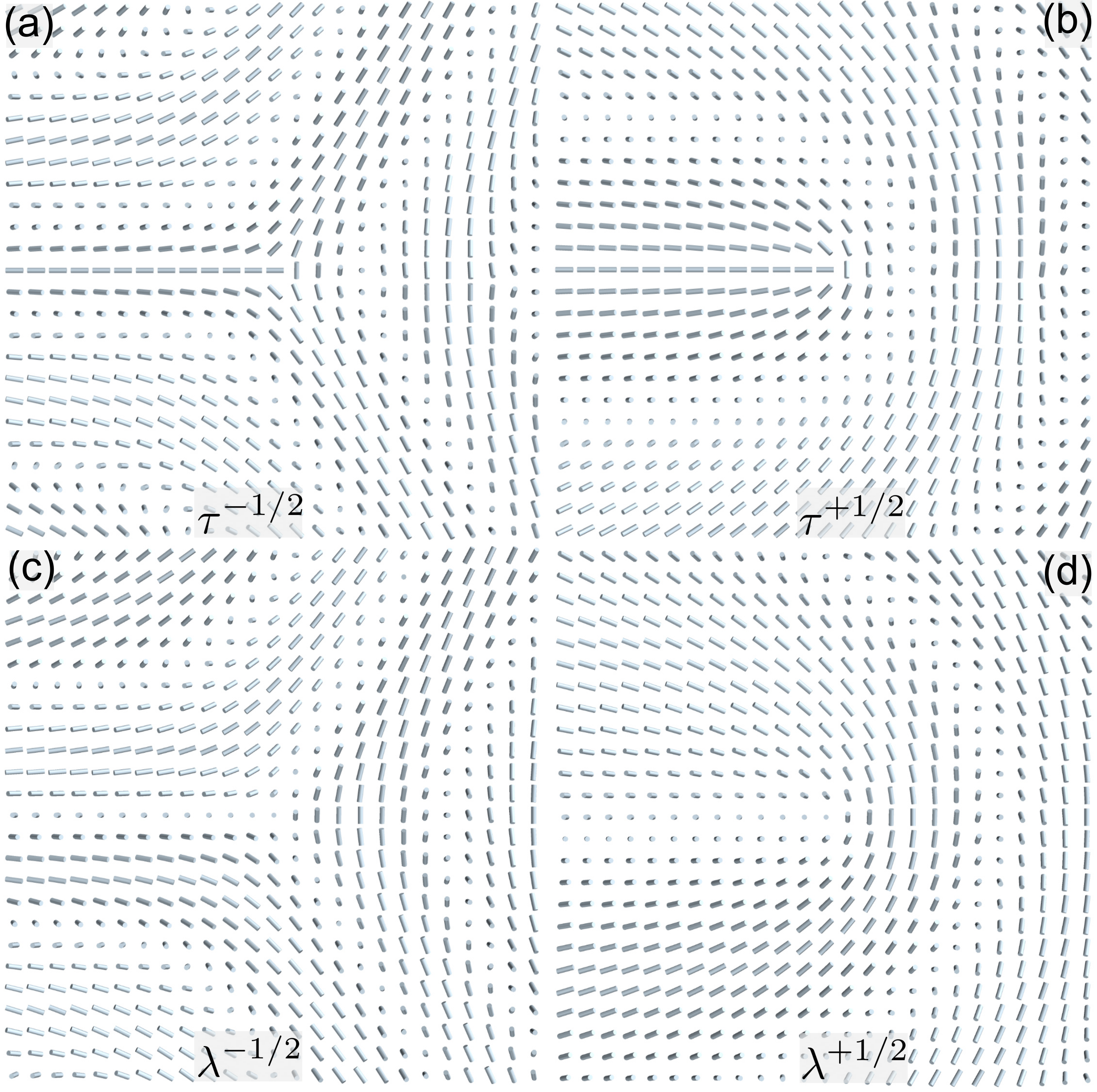}
    \caption{Examples of 2-dimensional $\tau$ and $\lambda$ disclination defect patterns. While $\tau$ defects (a,b) are singular in the director fields, $\lambda$ defects (c,d) are not. Defect profiles shown correspond to: $\tau^{-1/2}$ (a), $\tau^{+1/2}$ (b), $\lambda^{-1/2}$ (c) and $\lambda^{+1/2}$ (d).
    In 3D, the disclination line corresponding to these defect profiles is perpendicular to the plane of the drawing (or the $xy$ plane). The elements of Cl$(3,0,1)^{[0]}$ corresponding to the defect profiles are: $-e_{21}$ (a), $e_{21}$ (b), $-(\mathbb{1}+\frac{p}{4}e_{42})e_{13}e_{21}=-e_{32}+\frac{p}{4}e_{43}$ (c), and $e_{32}-\frac{p}{4}e_{43}$ (d).}
    \label{figtaulambda}
\end{figure}

The other two types of cholesteric defects, which are called $\tau$ and $\lambda$, both involve singularities in the helical axis so that the latter is not well defined near their core. Examples of such defects are the patterns shown in Fig.~\ref{figtaulambda}. The patterns corresponding to defects of type $\tau$ can be viewed as a simple rotation along an axis perpendicular to that of the screw (i.e., an axis perpendicular to the plane of the drawing in Fig.~\ref{figtaulambda}). Similarly, defects of type $\lambda$ can be viewed as the same rotation (represented by $e_{13}$ and $-e_{13}$ for $\pi$ and $-\pi$ rotations respectively) followed by a screw transformation along the helical axis (a direction in the plane of the drawing in Fig.~\ref{figtaulambda}) of size $p/2$. 
Therefore, these two types of defects can be algebraically represented as follows, 
\begin{eqnarray}\label{screwrep1}
    \tau & \leftrightarrow & e_{13} 
    \\ \nonumber
    \lambda & \leftrightarrow & -\Sigma_{z}\left(\frac{p}{2},p\right)e_{13}=\left(\mathbb{1}+\frac{p}{4}e_{43}\right)e_{32}.
\end{eqnarray}
Note that it is also common to refer to the representation of $\tau$ in Eq.~(\ref{screwrep1}) as $\tau^{+1/2}$, as the corresponding rotation is by $+\pi$. [Correspondingly, $\tau^{-1/2}$ is represented by $-e_{13}$ in this Clifford algebra notation.]
In terms of dual quaternions, the representation of the $\tau$, $\chi$ and $\lambda$ defects in Eq.~(\ref{screwrep1}) becomes
\begin{eqnarray}\label{choldefrep}
    \tau & \leftrightarrow & j 
    \\ \nonumber
    \chi & \leftrightarrow & \left(1- \frac{p}{4} \epsilon k\right)k=k+\epsilon \frac{p}{4} 
    \\ \nonumber
    \lambda & \leftrightarrow & \left(1-\frac{p}{4}\epsilon k\right)i=i-\epsilon \frac{p}{4}j,
\end{eqnarray}
where we have identified the dual unit $\epsilon$ (such that $\epsilon^2=0$) with $e_{4321}$, and the quaternions $(i,j,k)$ with the triplet of bivectors $(e_{32},e_{13},e_{21})$. 

It is useful to pause to discuss a few key properties of the representation in Eq.~(\ref{choldefrep}). 

First, the antidefects for $\tau$ and $\lambda$ are $-\tau$ and $-\lambda$. Therefore these two defect profiles can be viewed as Majorana spinors, as in the nematic case, because spinors are defined up to a minus sign. It can also be seen that both the spinors square to $-\mathbb{1}$ as in the nematic case. The case of the screw dislocation spinor is different though. This corresponds to a screw transformation, which does {\it not}  square to $-\mathbb{1}$, instead $\chi^2=-\mathbb{1}-\frac{p}{2}e_{43}=-1+\frac{p}{2}\epsilon k$, so that the square of $\chi$ entails a translation (of $\frac{p}{2}$) as well. As a consequence, generally $-\chi$ defects are {\it not} the inverse of $\chi$, and hence they cannot be be mathematically viewed as a Majorana spinor. For a point-like defect, we might use the fact that cholesteric patterns are invariant after translation of half a pitch (which are associated with $2\pi$ rotations), to neglect factors of $\frac{p}{2}e_{43}$ and hence reestablish a Majorana-like interpretation for $\chi$. However, this symmetry breaks down for extended disclinations (i.e., non-point-like defects, such as any segments in the lines in Fig.~\ref{figscrew}). For segments of size $l$, the relevant algebraic representation of defects as Clifford algebra elements is
\begin{eqnarray}\label{chirep}
    \chi & \leftrightarrow & \Sigma_z(p/2+l,p) \\ \nonumber &=&
    \left[\mathbb{1}+\frac{z}{2} e_{43}\right] \left[\cos\left(\frac{\phi}{2}\right)\mathbb{1} +\sin\left(\frac{\phi}{2}\right)e_{21}\right],
    \\ \nonumber 
    z & = & \frac{p}{2}+l, \qquad \phi = \frac{2\pi z}{p}.
\end{eqnarray}
Its inverse is now different from $-\chi$, even if we neglect translations of any multiple of half a pitch: this is because the inherent chirality of cholesterics breaks the symmetry between right-handed and left-handed local screw disclination profiles, such that the inverse of a $+1/2$ right-handed profile is a right-handed $-1/2$ profile, given explicitly by
\begin{equation}\label{chiinvrep}
\left[\mathbb{1}-\frac{z}{2} e_{43}\right] \left[\cos\left(\frac{\phi}{2}\right)\mathbb{1} -\sin\left(\frac{\phi}{2}\right)e_{21}\right],
\end{equation}
rather than the left-handed $-1/2$ profile which Eq.~(\ref{chirep}) transforms onto under a parity transformation.  Therefore, (extended, or non-point-like) $\chi$ defects are more naturally interpreted as Weyl-like, rather than Majorana-like, spinors, as they have a well-defined chirality, or equivalently correspond to a specific combination of rotations and translations. This is qualitatively similar to Weyl spinors in particle physics, which correspond to a specific combination of rotations and boosts. 

Second, the three elements corresponding to $\lambda$, $\tau$ and $\chi$, unlike the case of the biaxial nematics, do not provide a close subgroup. Indeed, it can be seen that $\lambda \tau =\chi$ and $\chi \lambda=\tau$, but $\tau \chi=\lambda -\frac{p}{2}e_{42}=\lambda +\epsilon \frac{p}{2}j$, hence the combination entails an extra translation. Again, this is non-negligible for non-point-like defects. We suggest that the neglect of this feature is the origin of the cholesteric algebra puzzle described in~\cite{beller2014} and reviewed in the Introduction.

Finally, we note that the $\lambda$ and $\tau$ defects are related by left multiplication by a screw transformation of size $\frac{p}{2}$ (a translation of $\frac{p}{2}$ and a rotation of $\pi$). Because of their spinor nature, this operation only rotates the patterns with respect to each other by $\frac{\pi}{2}$, and translates them by $\frac{p}{4}$, which corresponds to our physical intuition of the relation between the two defects in Fig.~\ref{figtaulambda}. [In contrast, a screw transformation $\Sigma$ on a Clifford vector $V$ would act via a sandwich product, or via $\Sigma V \Sigma^{-1}$.]

\begin{figure}[t!]
    \centering
    \includegraphics[width=0.48\textwidth]{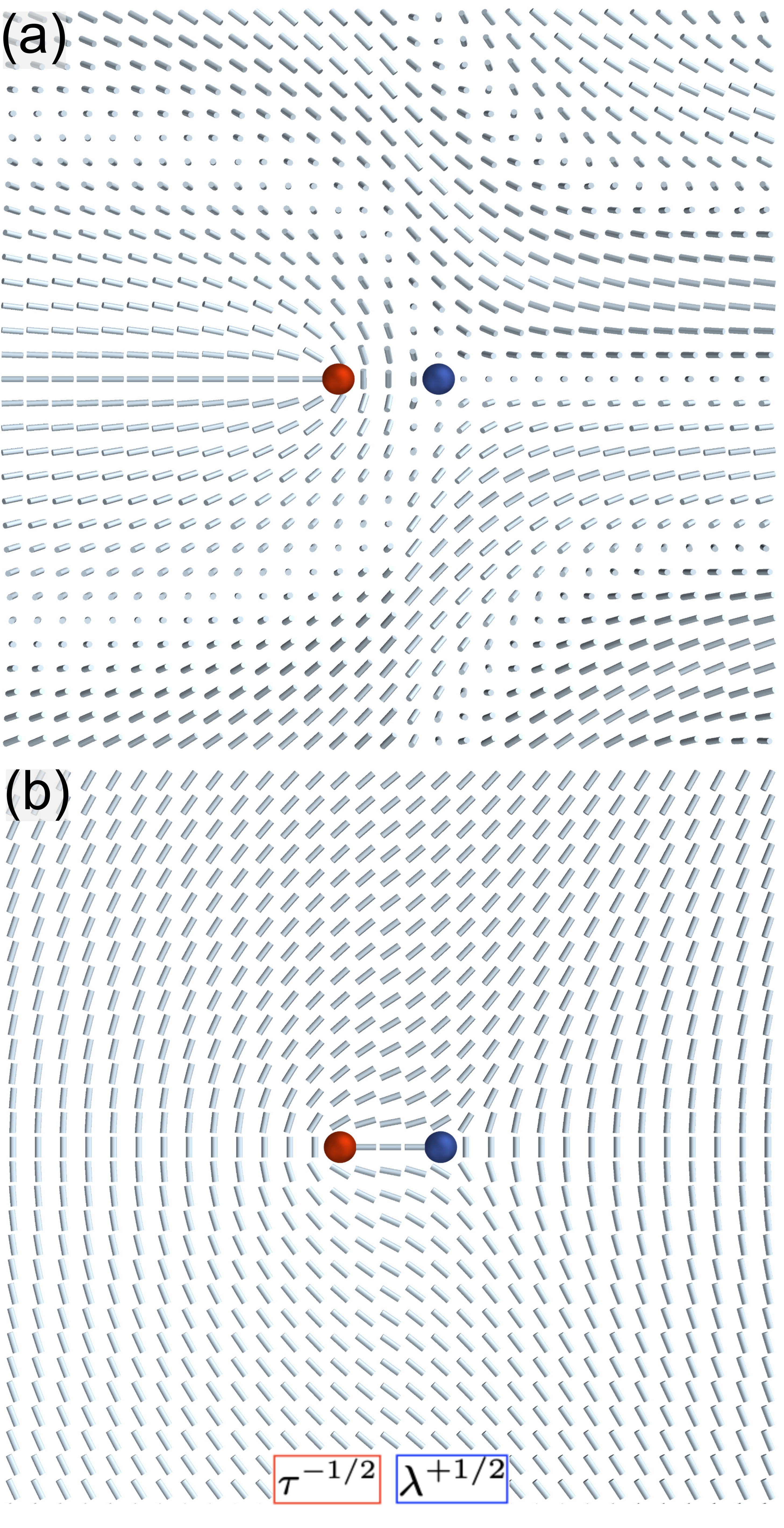}
    \caption{Example of edge dislocation pattern. (a) Director field profile, showing the splitting of an edge dislocation into $\tau$ and $\lambda$ defects, here a $\tau^{-1/2}$ and a $\lambda^{+1/2}$ pair. (b) Helical direction ($\chi$) pattern, showing singularities at the location of the $\lambda$ and $\tau$ defects, and the parallel far field along the vertical direction. } 
    \label{figedge}
\end{figure}

While in screw dislocations the disclination tangent ${\hat{\mathbf T}}$ is parallel to the helical axis $\hat{{\mathbf h}}$ (Fig.~\ref{figscrew}), another relevant geometry is the one in which  ${\hat{\mathbf T}}$ is perpendicular to $\hat{{\mathbf h}}$. In this geometry, the relevant $\chi$ dislocation is an edge dislocation, associated with a twist disclination, such that the Burgers vector is now perpendicular to the disclination (Fig.~\ref{figedge}). Taking ${\hat{\mathbf T}=e_3}$, along the $z$ direction, we can associate $\chi$ with a screw along the $y$ direction, $\tau$ with a rotation along the $z$ direction, and $\lambda$ again as a rotation along $z$ followed by a screw transformation along a perpendicular direction (here $y$). Algebraically, this corresponds to 
\begin{eqnarray}\label{edgerep1}
    \tau & \leftrightarrow & e_{21} 
    \\ \nonumber
    \chi & \leftrightarrow & \left(\mathbb{1}+\frac{p}{4} e_{42}\right)e_{13} = e_{13}+\frac{p}{4} e_{4321}
    \\ \nonumber
    \lambda & \leftrightarrow & -\left(\mathbb{1}+\frac{p}{4} e_{42}\right)e_{13}e_{21} = 
    e_{23}+\frac{p}{4} e_{43}.
\end{eqnarray}
In terms of dual quaternions, Eq.~(\ref{edgerep1}) becomes
\begin{eqnarray}\label{edgerep2}
    \tau & \leftrightarrow & k 
    \\ \nonumber
    \chi & \leftrightarrow & (1- \frac{p}{4} \epsilon j)j=j+\frac{p}{4} \epsilon 
    \\ \nonumber
    \lambda & \leftrightarrow & -(1-\frac{p}{4} \epsilon j)jk=-i- \frac{p}{4}\epsilon k.
\end{eqnarray}

Similar consideration holds for this representation, in Eqs.~(\ref{edgerep1},\ref{edgerep2}), as for the previous one, given by Eq.~(\ref{choldefrep}). In this geometry, this representation predicts that $\chi=\tau\lambda$, in line with the well-known fact that a $\chi$ edge dislocation can be seen as a composite defect~\cite{lavrentovich2001,smalyukh2002}, or a topological dipole made up, for instance, by a $\tau-\lambda$ pair (in the example in Fig.~\ref{figedge}). Here, the algebraic formula accounts for the $\frac{p}{4}$ effective size of the edge dislocation -- again the finite size renders this structure a Weyl-like spinor algebraically. The explicit pattern in Fig.~\ref{figedge} shows, in addition, that the $\chi$ field (the helical axis), as expected, is well defined in the far field, and everywhere except near the $\tau$ and $\lambda$ defects. This edge dislocation pattern is an example of composite quasiparticle, where two spinors ($\tau$ and $\lambda$) combine to form another spinor ($\chi$): the composite quasiparticle is therefore fermionic in nature here.

The magnitude of the Burgers vector in the edge dislocation in Fig.~\ref{figedge} is $b=\frac{p}{2}$. It is interesting to also consider the case of dislocation with $b=p$, which theory suggests to be a pair of $\lambda$ defects. These dislocations appear in thicker region of Grandjean-Cano wedges~\cite{smalyukh2002,thapa2024}, and are naturally found in ferroelectric chiral nematics~\cite{zhao2021,thapa2024}.

In the Clifford algebra formalism, a (finite-size) composite of $\lambda^{-1/2}$ and $\lambda^{+1/2}$ type elements in Eq.~(\ref{edgerep1}) can either equal $\mathbb{1}$, corresponding to a trivial defect-free pattern, or to $(\mathbb{1}\pm \frac{p}{2}e_{42})$, corresponding to a pure translation along the helical axis. Algebraically, which result is selected depends on the order in which the rotations and translations which make up each of the $\lambda$ elements are composed. For instance, if we choose
\begin{eqnarray}
\lambda_1 & \leftrightarrow & -\left(\mathbb{1}+\frac{p}{4} e_{42}\right)e_{13}e_{21}\\ \nonumber
    \lambda_2 & \leftrightarrow & e_{13}e_{21}\left(\mathbb{1}+\frac{p}{4} e_{42}\right),
\end{eqnarray}
we find
\begin{equation}\label{pdislocation}
    \lambda_1 \lambda_2 \leftrightarrow \left(\mathbb{1}+\frac{p}{2} e_{42}\right),
\end{equation}
which corresponds to an edge dislocation with $b=p$. [Instead, choosing $\lambda_2=-\lambda_1$ gives $\lambda_1\lambda_2=\mathbb{1}$.] In the case of $b=p$ dislocations, the resulting quasiparticles is not fermionic, as Eq.~(\ref{pdislocation}) does not square to $-\mathbb{1}$, but is instead bosonic in nature. Due to the presence of a non-trivial Burgers vector, the composite can be thought of as a vector boson.  Therefore the Clifford algebra formalism shows there is a fundamental algebraic difference between edge dislocations with $b=\frac{p}{2}$ and $b=p$.

\subsection*{Defect bivectors in cholesterics}

\begin{figure*}[t!]
    \centering
    \includegraphics[width=1.9\columnwidth]{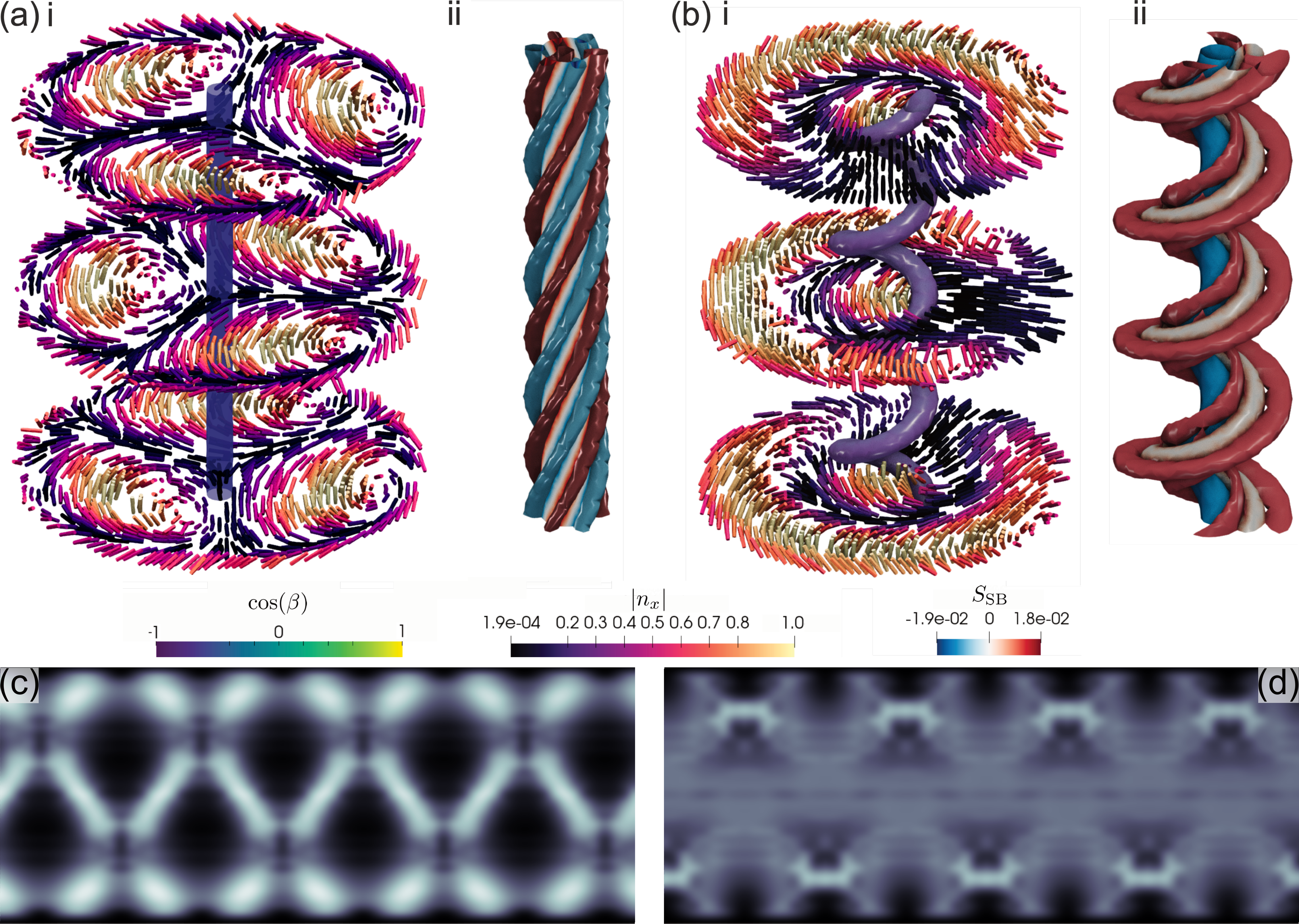}
    \caption{$\chi^{-1/2}$ (a) and  $\chi^{+1/2}$ (b) screw dislocation inside a cholesteric cylinder. Disclinations pierce the cylinder along the central axis. These are visualized as isosurfaces of the highest eigenvalue of the $Q$ tensor ($\lambda^\downarrow(Q)=0.1$) in (a)i and (b)i, and colored accordingly to $\cos(\beta)$. In both cases the director field $\mathbf{n}$ is visualized as rods on three planes cutting the principal axis of the cylinder, and colored accordingly to $\abs{n_x}$. Panels (a)ii and (b)ii show the isosurfaces of the splay-bend parameter $S_{SB}$ around the disclination lines (blue corresponding to $S_{SB}=-0.01$, while red to $S_{SB}=0.01$). (c-d) Simulated polarized optical microscopy images for corresponding to the configurations in (a) and (b), respectively (see Appendix for more details).} 
    \label{figconfinedchi}
\end{figure*}

Generally, defects in cholesterics have a rotational part (disclination) and a translational part (dislocation). Therefore, they can be described by the combination of a rotation of $\pi$ around a unit vector ${\mathbf{\Omega}}$ followed by a translation along a Burgers vector ${\mathbf b}$ (which has a variable length, corresponding to the extent of the disclination). Algebraically, we can therefore associate a local director profile at a defect with the following linear combination of bivectors
\begin{eqnarray}\label{defectbivectorchol}
    \Delta & = & \left[\mathbb{1}+{\mathbf b}\cdot{\mathbf e}_t\right]  
     {\mathbf \Omega}\cdot{\mathbf e}_r   
    \\ \nonumber & = & {\mathbf{\Omega}}\cdot {\mathbf e_{r}} +
    \epsilon {\mathbf{\Omega}}\cdot {\mathbf b} + 
    \epsilon \left({\mathbf{\Omega}}\times {\mathbf b}\right)\cdot {\mathbf e}_r \, ,
\end{eqnarray}
which can describe generic cholesteric defect patterns, and generalises the defect bivector introduced for nematics in Eq.~(\ref{defectbivector}).  
We suggest $\Delta$ can be used to quantify the local profile both geometrically and topologically. 
It is interesting to note that Eq.~(\ref{defectbivector}) is formally analogous to the representation of the Faraday bivector in the space-time Clifford algebra Cl(1,3)~\cite{hestenes2003,doran2003}.

The vector ${\mathbf{b}}$ is, for both screw and edge disclinations, parallel to the helical pitch axis $\hat{\mathbf{h}}$. Like ${\hat{\mathbf{\Omega}}}$, $\hat{\mathbf{h}}$ can be reconstructed starting from the ${\mathbf Q}$ tensor. Specifically, this can be done by using the chirality tensor, 
\begin{equation}
\label{defCij}    
C_{ij}=\epsilon_{j\mu\nu}Q_{\mu\alpha}\partial_{i}(Q_{\nu \alpha}).
\end{equation}
In the uniaxial limit, $Q_{\alpha\beta} \propto \left(n_{\alpha}n_{\beta}-\delta_{\alpha\beta}/3\right)$, this tensor can be written in terms of the director field as $C_{ij}=\epsilon_{j\mu\nu}n_{\mu}\partial_{i}n_{\nu}$, which is the chirality tensor defined in~\cite{beller2014}. The helical axis can be found as the left eigenvector of $C_{ij}$, which can be identified by singular value decomposition. The trace of $C_{ij}$ can be written as 
\begin{equation}\label{TrC}
{\rm Tr}(C)=-\epsilon_{\alpha \gamma \delta}Q_{\alpha\beta}\partial_{\gamma}Q_{\delta\beta}=-S_{\rm tw},
\end{equation}
where $S_{\rm tw}$ is the twist parameter defined in~\cite{Fialho2017,Copar2013_visualization,KilianSonnet}.

\subsection*{Matrix representation and Cayley factorisation of cholesteric spinors}

Until now, we have used Cl(3,0,1) elements without any reference to a specific representation. It is also useful in practice to provide a suitable matrix representation both of generic elements of the Clifford algebra and of the even subalgebra (again in terms of matrices, rather than dual quaternions). Elements in this Clifford algebra can be 
represented by a subset of $4\times 4$ matrices, with the following generators,
\begin{eqnarray}
    e_1 =
    \begin{bNiceMatrix}
        0 & 0 & 0 & +i \\
        0 & 0 & +i & 0 \\
        0 & -i & 0 & 0 \\
        -i & 0 & 0 & 0
    \end{bNiceMatrix} \,
    e_2 = 
    \begin{bNiceMatrix}
        0 & 0 & 0 & -1 \\
        0 & 0 & +1 & 0 \\
        0 & +1 & 0 & 0  \\
        -1 & 0 & 0 & 0
    \end{bNiceMatrix} \\ \nonumber
    e_3 = 
    \begin{bNiceMatrix}
        0 & 0 & -i & 0 \\
        0 & 0 & 0 & +i \\
        +i & 0 & 0 & 0 \\
        0 & -i & 0 & 0
    \end{bNiceMatrix} \,
    e_4 = 
    \begin{bNiceMatrix}
        +i & 0 & +1 & 0 \\
        0 & +i & 0 & +1 \\
        +1 & 0 & -i & 0 \\
        0 & +1 & 0 & -i
    \end{bNiceMatrix}
\end{eqnarray}
where $e_1^2=e_2^2=e_3^2=1$, $e_4^2=0$, and $e_i e_j=-e_je_i$ for $i\ne j$.




As cholesteric spinors are combinations of even-grade elements in Cl(3,0,1)$^{[0]}$, it is useful to have a representation of this even subalgebra, which can be achieved by exploiting Cayley factorisation of $4\times 4$ matrices into right and left isoclinic bases~\cite{perez2017,hunt2016}. A right and a left basis set is provided by the following $A_{x,y,z}$ and $B_{x,y,z}$ matrices respectively,
\begin{eqnarray}\label{rotrep}
     A_{x} = \begin{bNiceMatrix}[r][columns-width=auto]
         0 &  0 &  0 & -1 \\
         0 &  0 & -1 &  0 \\
         0 & +1 &  0 &  0 \\
        +1 &  0 &  0 &  0
    \end{bNiceMatrix} \,
    B_{x} = \begin{bNiceMatrix}[r][columns-width=auto]
         0 &  0 &  0 & +1 \\
         0 &  0 & -1 &  0 \\
         0 & +1 &  0 &  0 \\
        -1 &  0 &  0 &  0
    \end{bNiceMatrix} \\ \nonumber
     A_{y} = \begin{bNiceMatrix}[r][columns-width=auto]
         0 &  0 & +1 &  0 \\
         0 &  0 &  0 & -1 \\
        -1 &  0 &  0 &  0 \\
         0 & +1 &  0 &  0
    \end{bNiceMatrix} \,
     B_{y} = \begin{bNiceMatrix}[r][columns-width=auto]
         0 &  0 & +1 &  0 \\
         0 &  0 &  0 & +1 \\
        -1 &  0 &  0 &  0 \\
         0 & -1 &  0 &  0
    \end{bNiceMatrix} \\ \nonumber
     A_{z} = \begin{bNiceMatrix}[r][columns-width=auto]
         0 & -1 &  0 &  0 \\
        +1 &  0 &  0 &  0 \\
         0 &  0 &  0 & -1 \\
         0 &  0 & +1 &  0
    \end{bNiceMatrix} \,
     B_{z} = \begin{bNiceMatrix}[r][columns-width=auto]
         0 & -1 &  0 &  0 \\
        +1 &  0 &  0 &  0 \\
         0 &  0 &  0 & +1 \\
         0 &  0 & -1 &  0
    \end{bNiceMatrix}
\end{eqnarray}
It can be shown that the matrix algebra of $(A_{x},A_{y},A_{z})$ [or of $(B_{x},B_{y},B_{z})$] is that of quaternions, hence the $A$ and $B$ triads can both be mapped to the triad of bivectors $(e_{32},e_{13},e_{21})$. Only one set of matrices is needed to represent 3D transformations with sandwich products.

While the matrices in Eq.~(\ref{rotrep}) describe rotations, translations can be represented by $(\epsilon A_{x},\epsilon A_{y},\epsilon A_{z})$, or by $(\epsilon B_{x},\epsilon B_{y},\epsilon B_{z})$, which give alternative basis sets of isoclines. 

A general element of Cl(3,0,1)$^{[0]}$ -- or a unit dual quaternion corresponding to a roto-translation -- can thus be written in terms of isoclinic rotation matrices as follows (taking, for instance, the $A_{x,y,z}$ basis),
\begin{equation}\label{SigmaCayley}
    \Sigma =  \left[\mathbb 1 -\epsilon \frac{d {\mathbf b}\cdot{\mathbf A}}{2}\right]\left[\cos(\phi) \mathbb{1}+ \sin(\phi)  {\mathbf a}\cdot{\mathbf A}\right], 
\end{equation}
where ${\mathbf{a}}$ and ${\mathbf{b}}$ are the unit vectors describing the translation and rotation, as in Eq.~(\ref{rototranslation}), to which Eq.~(\ref{SigmaCayley}) is equivalent. 
To see this, we only need to recall that the $A$ basis can be identified with
$(e_{32},e_{13},e_{21})$, and that $\epsilon$ can be identified with $e_{4321}$. 


An advantage of the formalism described in this section is that the corresponding representation of the even subalgebra Cl(3,0,1)$^{[0]}$, and hence of cholesteric spinors, is real and constitutes a Lie algebra for $e_{32}$, $e_{13}$, and $e_{21}$. The representation of the translations, whilst requiring the dual number $\epsilon$, is also a natural generalisation of this algebra. 

\subsection*{Geometry of screw dislocations under cylindrical confinement}
\begin{figure*}[t!]
    \centering
    \includegraphics[width=2.0\columnwidth]{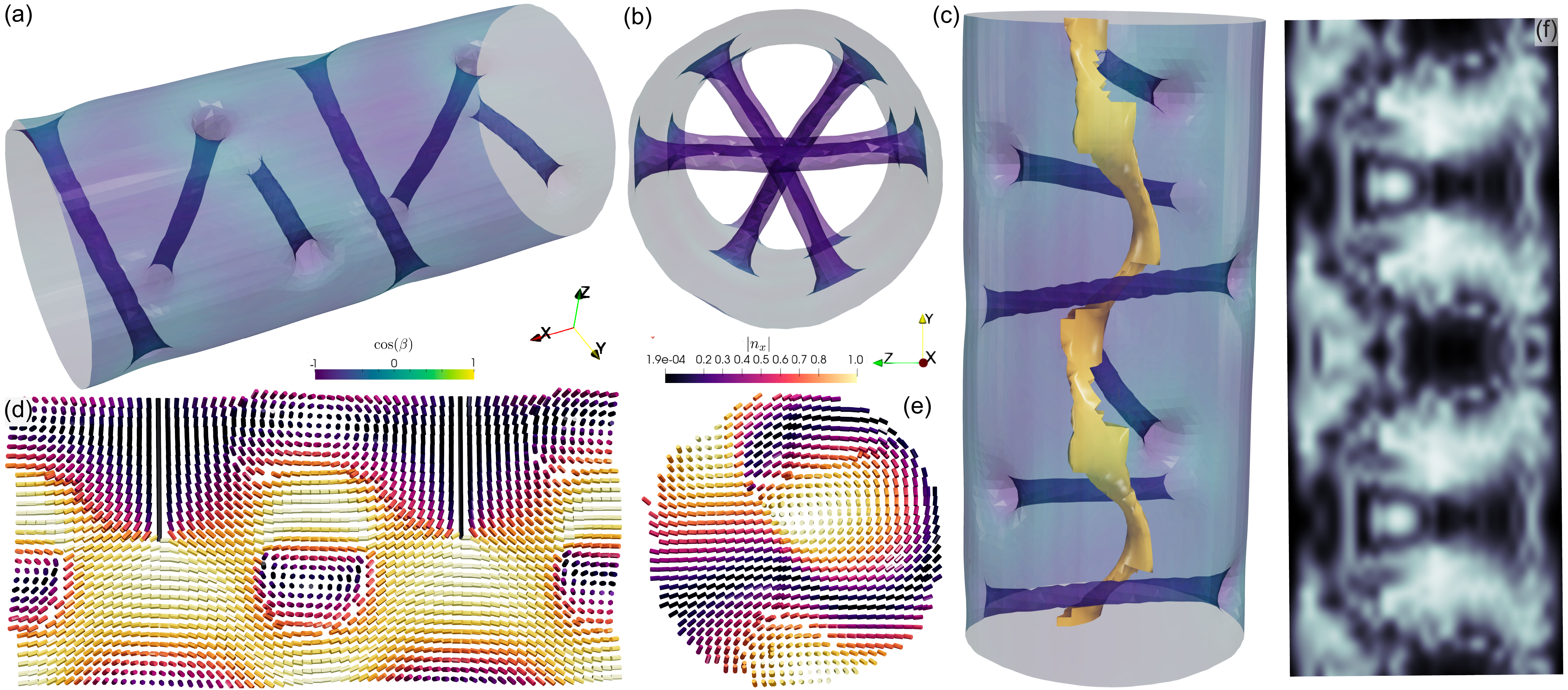}
    \caption{(a,b,c) Side (a) and top (b) view of a cartwheel disclination pattern. In (c) the region where the director field is aligned along the axis $\hat{\mathbf x}$ of the cylinder (with projection ${\mathbf{n}}\cdot \hat{\mathbf{x}}\ge 0.99$) is also shown, to visualise skyrmion filaments along the cylindrical axis. This cartwheel structure results from an initialisation with a straight disclination along the cylinder axis, and a spiralling $\tau^{-1/2}$ defect profile on the cross-section (see Appendix for initialisation details and parameter list). (d,e) Cross section along two planes, which are respectively parallel (d) and perpendicular (e) to the cylinder axis, of the corresponding steady-state director field pattern. (f) Simulated polarized optical microscopy image.}
    \label{figrotatingtau}
\end{figure*}

The algebraic representation discussed above provides a powerful way to predict results from combinations of defects. In some cases, the predicted patterns can be observed in practice, such as for the composite edge dislocation shown in Fig.~\ref{figedge}, which spontaneously form in a cholesteric wedge~\cite{smalyukh2002}. However, in general an algebraically plausible pattern need not be realised in practice, if, for instance, it costs loo large an amount of elastic free energy. 

To explore the thermodynamic stability of cholesteric spinors and defects, in this Section we focus on the geometry and structure of screw dislocation patterns, which are comparatively less studied with respect to edge dislocations. Specifically, we analyse configurations found numerically under cylindrical confinement, when initialising the system with a defect pattern corresponding to a screw dislocation, as in Fig.~\ref{figscrew} -- in each cross section perpendicular to the disclination, this structure corresponds to a Weyl-like cholesteric spinor. Confinement is realised by choosing spatially dependent free energy parameters such that the liquid crystal is in the cholesteric phase inside a cylinder of radius $R$, and in the isotropic phase in the region outside the cylinder (see Appendix\ref{numappendix} for more details on the functional form of the free energy, the initialisation of the system and the choice of parameters). Our choice of cylindrical confinement is inspired by classical experiments in cholesteric tubes~\cite{kitzerow1996,lequeux1988,bouligand1974}, although in our cases the boundary and initial conditions are distinct and have been chosen to focus on the structure of screw dislocations.

Fig.~\ref{figconfinedchi} shows the results of simulations initialised with a $\chi^{-1/2}$ [Fig.~\ref{figconfinedchi}(a)] and with a $\chi^{+1/2}$ [Fig.~\ref{figconfinedchi}(b)] screw dislocation inside the cholesteric cylinder -- in other words, the starting configurations are those in Fig.~\ref{figscrew}. For the $\chi^{-1/2}$ case, this topology is stable, bar the appearance of double twist regions (skyrmions) surrounding the central triradius [Fig.~\ref{figconfinedchi}(a)]. In each of the cylindrical cross section, the director field pattern in the centre retains its Weyl-like character, where the sense of defect profile rotation matches the right-hand thermodynamic chirality of the underlying cholesteric phase. The disclination line has the topology of a twisted ribbon, as can be appreciated by visualising the splay-bend parameter -- defined as 
\begin{equation}
S_{\rm SB}=\partial_{\alpha}\partial_{\beta} Q_{\alpha\beta},
\end{equation}
which displays a ``barber-pole'' pattern in the director profile  close to the defect line [Fig.~\ref{figconfinedchi})(a)ii]. 

When starting from a $\chi^{+1/2}$ line, instead, the initially straight disclination curls in 3D to form a right-handed helix, where the cross section is a $-1/2$ profile which does not twist -- as can be seen by tracing the red splay filaments in the splay-bend pattern.  In other words, 
the inherent twist in the $\chi^{+1/2}$ line has transformed into 3D writhe. This morphological transition is similar to the conversion between twist and writhe which can be observed in supercoiled polymers, such as DNA, and that is mathematically described by the linking number theorem~\cite{BatesMaxell}. In our structure, a central double twist region, corresponding to a $\lambda$ line, accompanies the helical defect, such that the defect itself can be seen as an algebraically allowed $\lambda-\tau$ composite, similar to the case of the edge dislocation in Fig.~\ref{figedge} -- although here the composition is of the form $\chi^{+1/2} \leftrightarrow\tau^{-1/2}\lambda^{+1/2}\lambda^{+1/2}=\tau^{-1/2}\lambda^{+1}$. The composite nature of the defect can be visualised by tracing the splay-bend pattern in 3D [Fig.~\ref{figconfinedchi}(b)ii], as, given the defect geometry, splay and bend tend to localise close to $\tau$ and $\lambda$ defects respectively.

To ease comparison of our simulations with future experiments, we also provide simulated polarized optical microscopy images corresponding to the resulting screw dislocation patterns in Figs.~\ref{figconfinedchi}(c,d) (see Appendix and~\cite{chen2024}). When simulating these optical images, we assume that the cholesteric pitch is much larger than the wavelength of probing light, which is in accord  with the concept of ``weakly'' distorted cholesteric and smooth deformations of the director field.

Fig.~\ref{figrotatingtau} shows the results obtained when starting from a cholesteric $\tau^{-1/2}$ disclination spiralling along the $z$ direction (as in Fig.~\ref{figscrew}(a), but with a $\tau^{-1/2}$ pattern as in Fig.~\ref{figtaulambda})(a) instead of a triradius pattern at each plane). {\it A priori}, one may expect this to lead to the same pattern as for the rotating $\chi^{-1/2}$ disclination, as the patterns in the initial condition are quite similar. Instead, and surprisingly, numerical simulations show that the texture completely changes, and transforms into the 3D cartwheel pattern shown in Fig.~\ref{figrotatingtau}(a,b). Here, a set of $\tau$ disclinations perpendicular to the cylindrical axis is stabilised by a combination of bend deformations and an array of double twist cylinders (along mutually perpendicular directions). In this case, therefore, a single dislocation has broken up into a whole array of disclinations. 

\subsection*{Bosonisation and topological phases of cholesteric spinors}
\begin{figure*}[t!h]
    \centering
    \includegraphics[width=1.93\columnwidth]{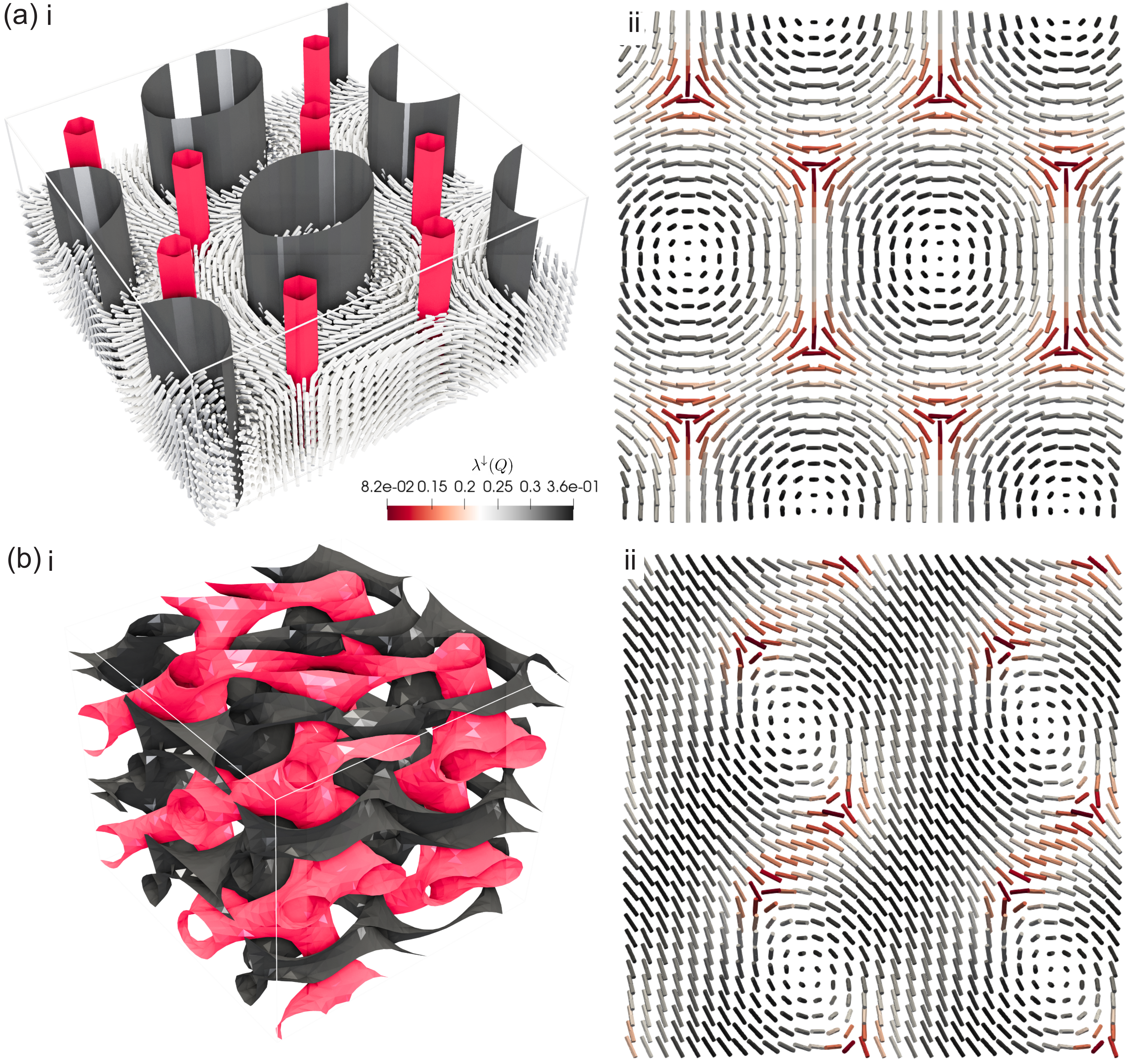}
    \caption{(a) Quasi-2D phase. (i) Skyrmion filaments (black), visualized as isosurfaces of the twist parameter $S_{\rm tw}=0.001$, and disclination lines (red) visualized as isosurfaces of the highest eigenvalue of the $Q$ tensor ($\lambda^\downarrow(Q)=0.1$). (ii) Cut showing the director field coloured accordingly to $\lambda^\downarrow(Q)$. 
    (b) 3D phase. (i) Skyrmion filaments and disclination lines in 3D. (ii) Cut showing the director field in the midplane.  Colours are as in (a).}
    \label{figskyrmions}
\end{figure*}

The striking morphological diversity of the patterns observed for confined screw dislocations resulting from different initial conditions, and discussed in the previous section, is reminiscent of the spontaneous appearance of blue phases in 3D bulk cholesterics, where a panoply of different metastable topological phases are found for the same thermodynamic parameters, close to the isotropic-cholesteric transition. As we shall now show, this phenomenon can be thought of as a manifestation of spinor condensation, or bosonisation. Here, again $\tau$ and $\lambda$ defects -- or skyrmion filaments -- combine to form composite quasiparticles. This time, they have bosonic nature, and can therefore condense to form topological phases. We highlight here, as well, that the formation of composite defects, which is at the basis of the condensation phenomenon, requires disclinations to entangle, or braid, with one another, which is only possible if the underlying spinor subalgebra is non-Abelian.



Because both $\lambda$ and $\tau$ disclinations have a topological charge, when they combine they can give rise to 
structures that are not topologically charged overall, as the half skyrmion-defect composites which are shown in Fig.~\ref{figskyrmions}. Because their effective overall topological charge is $0$, these defect-skyrmion complexes behave as composite bosons, and can tile the plane to form 2D hexagonal lattices (Fig.~\ref{figscrew}(a)i,ii; these two-dimensional blue phases are only thermodynamically stable under an electric field~\cite{hornreich1985,henrich2010}).  With respect to the case of edge dislocations with $b=p$, these bosonic excitations have scalar, rather than vectorial, character algebraically. These composite quasiparticles are also qualitatively similar to Cooper pairs in superconductors, which are composite bosons made by pairs of fermions~\cite{bardeen1957}. 

To gain more insight into cholesteric spinor condensation, or bosonisation, phenomenon, we numerically study a periodic system with periodic boundary conditions and variable thickness. To visualise skyrmions, we have used the local twist parameter $S_{\rm tw}$ defined in Eq.(~\ref{TrC})~\cite{Fialho2017}.
Specifically, we have identified skyrmion filaments with regions in space where the absolute value of $S_{\rm tw}$ is above a given threshold. Physically, this is where twist is maximum, which generalises the notion of double twist regions in 2D [Fig.~\ref{figskyrmions}(a)].

For thin samples, the hexagonal 2D blue phase is stable, and it stretches into the third dimension to form an array of cylindrical skyrmions separated by straight $-1/2$ disclination lines [Fig.~\ref{figskyrmions}(a)]. This structure is also the one found for thin samples in  Refs.~\cite{pisljar2022,pisljar2023}. 

Activity~\cite{metselaar2019} and geometry~\cite{carenza2022} were found to morph these ordered lattices into amorphous or quasicrystalline structures. Here we focus on the effect of sample thickness on the morphology of the disclination line patterns. For sufficiently thick samples, we find that the quasi-2D hexagonal lattice is destabilised and gives way to a fully-3D network of disclination lines, which is interpenetrated by an analogous network of skyrmion filaments [Fig.~\ref{figskyrmions}(b)], analogous to blue phases~\cite{pisljar2022,pisljar2023}.  The liquid crystal patterns shown in Fig.~\ref{figskyrmions} (as well as those reported in~\cite{pisljar2022,pisljar2023,carenza2022,metselaar2019}) are all examples of spinor condensation, as there is a finite density of defects in equilibrium, such that the number of defects is extensive in the cross-sectional area of the sample. For the cases considered in this Sections, all patterns, whether quasi-2D or fully-3D, are locally Majorana-like, as the defect profile along the disclination (away from the disclination junction points in Fig.~\ref{figskyrmions}(b)) have locally a triradius structure ($\hat{\mathbf \Omega}\cdot \hat{\mathbf T}=\cos(\beta)=-1$) throughout, and there is no twist of the cross-sectional defect profile along the disclination lines. The triradius feature is promoted thermodynamically, and it would be of interest to see whether and how activity changes it. Indeed, in active nematics, extensile, but not contractile, activity promotes twist-like defect profiles over triradii and comets~\cite{binysh2020,head2024b,adriano2024}.

\subsection*{Discussion and conclusions}

In this work, we have shown that local defect profiles in nematics and cholesterics behave as spinors, and as topological fermionic quasiparticles. For nematics, local 2D defect profiles are naturally interpreted as Majorana-like spinors, as the defects are topologically equivalent to the anti-defects which annihilate with them~\cite{head2024b}. Instead, cholesteric defects can be either Majorana-like (e.g., straight $\lambda$ and $\tau$ disclinations) or Weyl-like (e.g., extended $\chi$ dislocations). In the latter case, the chirality is selected thermodynamically, as the underlying free energy favours one chirality over the other, ultimately due to the microscopic shape of the molecules, or interactions between them.

These topological quasiparticles and their ``scattering properties'' after collisions are described by Clifford algebras, and we have seen that the algebras that describe different types of liquid crystal defects -- uniaxial nematics, biaxial, and cholesterics -- are different. 
We started by showing that the two-dimensional algebra Cl(2,0), and in particular its even subalgebra, which can be identified with the complex numbers, is sufficient to describe planar defects in uniaxial nematics. Instead, Cl(3,0) or the quaternion algebra is required to algebraically represent out-of-plane uniaxial defects. Biaxial defects can be described by two copies of Cl(2,0) or Cl(3,0), for planar or out-of-plane configurations respectively, where the two copies of the Clifford algebra correspond to the primary and secondary director field of the biaxial sample. Cholesterics require an even larger algebra than Cl(3,0), which accounts for both rotations and translations. This is Cl(3,0,1), the even subalgebra of which is equivalent to that of dual quaternions. The difference in the underlying algebra shows why the multiplication rules of quaternions apply to the composition of biaxial nematic defects but not always to that of cholesteric ones, as discussed in~\cite{beller2014}: this is because simple quaternions do not include translations along the helical axis as a physical transformation to consider.

The dual quaternion algebra allows us to understand how $\lambda$ and $\tau$ defects can combine to form edge dislocations, or helical screw $\chi$ dislocations under cylindrical confinement. Algebraically, these dislocations are composite defects with a fermionic nature, and accordingly the director field features a $\pi$ (rather than $2\pi$) rotation when we consider a circular path encircling the defect. 
Composites can also have a bosonic nature. An example is edge dislocations with Burgers vector equal to the pitch, which arise for instance in ferroelectric chiral nematics.
Another example is provided by topological phases emerging in 3D bulk cholesteric samples, such as blue phases or other skyrmion lattices, which are made up of unit cells which are topologically uncharged. These unit cells behave as composite quasiparticles with bosonic nature, and hence can condense to form space-filling lattices of disclinations. This phenomenon is reminiscent of the condensation of Cooper pairs in superconductors~\cite{bardeen1957}. 

The algebraic framework we have described also provide a natural way to describe cholesteric defects as a combination of disclinations and dislocations, as the $6$ bivectors in Cl(3,0,1)$^{[0]}$ account for the $3$ rotational and $3$ translational degrees of freedom identifying the Volterra process associated with the defect. This defect bivector constitutes a quantity akin to the electromagnetic tensor in quantum electrodynamics, and we hope it may provide a useful way to quantify the behaviour of cholesteric disclinations in 3D in the future. It would also be of interest to combine this algebraic description of cholesteric defects with that provided by contact topology in~\cite{pollard2023}.


As discussed in~\cite{head2024b}, a 2D local profile behaves like a quasiparticle at rest. Here, we have considered either 2D systems, where this analogy holds, or 3D systems where disclination lines can be viewed as extended quasiparticles, or quasi-strings. In 3D, a closer qualitative equivalent of a localised quasiparticle with finite momentum would be a loop, and it would be interesting to use our Clifford algebra framework to identify the nature of these types of quasiparticles in cholesterics and biaxial nematics. 

In conclusions, the theory we have outlined provides an unexpected connection between liquid crystal physics, particle physics and quantum condensed matter. It may provide novel ways to realise quantum mechanical behaviour at the classical scale, facilitating the study -- by experiments or computer simulations -- of the properties of and interactions between quasiparticles with different spinor natures. Looking ahead, 
activity provides an additional intriguing parameter to consider, as it constitutes a way to make these quasiparticles mobile~\cite{head2024,head2024b}, endowing them with nontrivial dispersion relations. 

\begin{acknowledgments}
This work used the ARCHER2 UK National Supercomputing Service (https://www.archer2.ac.uk), and was supported by US National Science Foundation grant DMR-2341830.

For the purpose of open access, the author has applied a Creative Commons Attribution (CC-BY) license to any Author Accepted Manuscript version arising from this submission.
\end{acknowledgments}

\appendix
\subsection*{Numerical simulations}\label{numappendix}
To model cholesteric liquid crystals in bulk and under cylindrical confinement we considered as dynamical fields the nematic $Q-$tensor ${\bm Q}({\bm r},t)$, whose principle eigenvector $\bm{n}$ --the so-called \emph{director field}-- defines the local direction of alignment of the liquid crystal (LC), and the incompressible velocity field $\bm{v}({\bm r},t)$.
The equilibrium properties of the system are described by the free energy
\begin{equation}
    \mathcal{F} = \int  \  \left ( f^{\rm bulk } + f^{\rm el }\right ) \,\ d\bm{r} \; , 
    \label{eqn:1}
\end{equation}
 where
\begin{gather}
\begin{split}
f^{\rm bulk } = A_0 \left[ \dfrac{1}{2} \left( 1-\frac{\chi}{3} \right) {\rm Tr} {\bm Q}^2  -\frac{\chi}{3} {\rm Tr} {\bm Q}^3 \right. \\ \left. +\frac{\chi}{4} \left( {\rm Tr} {\bm Q}^2 \right)^2 \right] \; ,
\end{split}  \label{eqn:3}  \\
f^{\rm el} = f^{\rm sb} + f^{\rm tw} = \frac{L}{2}\left[ (\nabla \cdot \bm{Q})^2 + (\nabla \times \bm{Q} + 2q_0 \bm{Q} )^2 \right]\; \label{eqn:4} . 
\end{gather}

In Eq.~\eqref{eqn:3}, the bulk constant $A_0>0$, while $\chi$ is a temperature-like parameter that drives the isotropic-nematic transition that occurs for $\chi>\chi_{cr}=2.7$~\cite{de1993physics}. In the case of cylindrical confinement, we set $\chi>\chi_{cr}$ inside the cylinder and $\chi<\chi_{cr}$ otherwise. The elastic free energy density in Eq.~\eqref{eqn:4}, proportional to the elastic constant $L$, captures the energy cost of elastic deformations in the single elastic constant approximation~\cite{de1993physics}. It has been split into a splay-bend contribution $f^{\rm sb}$ and a twist contribution $f^{\rm tw}$, which stabilizes helical structures when the chiral wavenumber $q_0$ is non-zero. For $q_{0}>0$, the equilibrium configuration in unconfined geometries features a right-handed helix with pitch $p_0=2\pi/q_0$.

The following set of coupled PDEs governs the dynamics of the fields,
\begin{eqnarray}
&D_t{\bm Q}= {\bm S}({\bm W},{\bm Q}) + \gamma^{-1} {\bm H} \; , \label{eqn:7} \\
&\rho D_t {\bf v} = \nabla\cdot ({\bm \sigma}^{\rm hydro}+{\bm \sigma}^{\rm LC}) \; , \label{eqn:8} 
\end{eqnarray}
where the differential operator $D_t= \partial_t + \bm{v}\cdot \nabla$ is the material derivative.
Eq.~\eqref{eqn:7} is the Beris-Edwards equation ruling the dynamics of the $Q-$tensor. The operator $\bm{S}(\bm{W},\bm{Q})$ denotes the co-rotational derivative and defines the dynamical response of the LC to straining and shearing. Its explicit expression depends on both the velocity gradient $\bm{W} = \nabla \bm{v} $ and the $Q-$tensor configuration (see Eq.~\ref{eqnA1} for the explicit expression). 
The coefficient $\gamma$ is the rotational viscosity measuring the importance of advection relative to relaxation, and the molecular field ${\bm H}=-\frac{\delta {\mathcal F}}{\delta {\bm Q}}+({\bm I}/3){\rm Tr}\frac{\delta {\mathcal F}}{\delta {\bm Q}}$.
Finally, Eq.~\eqref{eqn:8} is the Navier-Stokes equation for the incompressible velocity field ($\nabla \cdot \bm{v} = 0$) with constant density $\rho$. Here, the stress tensor has been divided in: \emph{(i)} a hydrodynamic contribution ${\bm \sigma}^{\rm hydro}= - P \bm{I} + \eta \nabla \bm{v}$ accounting for the hydrodynamic pressure $P$ ensuring incompressibility, and viscous effects, proportional to the viscosity $\eta$ and \emph{(ii)} a LC contribution ${\bm \sigma}^{\rm LC}$ accounting for elastic and flow-aligning effects.
The explicit expression of the corototational derivative $\bm{S}({\bm W},{\bm Q})$ appearing in Eq.~\eqref{eqn:7} is given by
\begin{eqnarray}
\bm{S}({\bm W},{\bm Q}) & = & (\xi{\bm D}+{\bm \Omega})({\bm Q}+{\bm I}/3)\\
& + & (\xi{\bm D}-{\bm\Omega})({\bm Q}+{\bm I}/3) \\
& - & 2\xi({\bm Q}+{\bm I}/3) {\rm Tr} ({\bm Q}{\bm W}). 
\label{eqnA1}
\end{eqnarray}
Here, ${\bm D}=({\bm W}+{\bm W}^T)/2$ and ${\bm\Omega}=({\bm W}-{\bm W}^T)/2$ are the symmetric and anti-symmetric part of the velocity gradient tensor $W_{\alpha\beta}=\partial_{\beta}v_{\alpha}$, respectively. The flow alignment parameter $\xi$ determines the aspect ratio of the LC molecules and the dynamical response of the LC to an imposed shear flow. Here, we choose $\xi = 0.7$ to consider flow-aligning rod-like molecules. 

The explicit expression of the LC contribution is 
\begin{eqnarray}
\sigma_{\alpha\beta}^{LC}=&&-\xi H_{\alpha\gamma}(Q_{\gamma\beta}+\frac{1}{3}\delta_{\gamma\beta})-\xi(Q_{\alpha\gamma}+\frac{1}{3}\delta_{\alpha\gamma})H_{\gamma\beta}\nonumber\\
&&+2\xi(Q_{\alpha\beta}+\frac{1}{3}\delta_{\alpha\beta})Q_{\gamma\mu}H_{\gamma\mu}+Q_{\alpha\gamma}H_{\gamma\beta}-H_{\alpha\gamma}Q_{\gamma\beta}\nonumber\\
&&-\partial_{\alpha}Q_{\gamma\mu}\frac{\partial f}{\partial(\partial_{\beta}Q_{\gamma\mu})}.
\end{eqnarray}

We integrated the dynamics of the hydrodynamic fields in Eq.~\eqref{eqn:8} in a cubic grid of size $\mathcal{L}=32,64,128$ using a predictor-corrector hybrid lattice Boltzmann approach~\cite{succi2018,carenzaepje}. This consists of solving Eq.~\eqref{eqn:7}  with a finite-difference algorithm implementing the first-order upwind scheme and fourth-order accurate stencils for space derivatives, and the Navier-Stokes equation through a predictor-corrector LB scheme on a $D3Q15$ lattice. For technical details on the lattice Boltzmann method see~\cite{carenzaepje,Negro2024}.


In all simulations the velocity field was initialised to $0$, while the ${\mathbf Q}$ tensor was initially set as described in the two following sections. These sections also contain the full parameter sets used to obtain the configurations in Figures 6-8.

\subsubsection*{Cholesteric liquid crystals under cylindrical confinement}

The results presented in Figs. 6 and 7, were obtained by confining a cholesteric liquid crystal inside a cylinder (with no, or free, anchoring at the boundaries). This has been achieved by imposing $\chi=3$ (corresponding to the cholesteric phase) inside a cylinder of radius $R=15$, and $\chi=2$ (corresponding to the isotropic phase) otherwise. Other parameters were set as follows: $q_0=\pi/16$, $A_0=0.02$, $L=0.01$, $\xi=0.7$, $\gamma=0.33775$ and the viscosity $\eta=5/3$. The simulation box was a cube with size $L_x=L_y=L_z=64$ and periodic boundary conditions. The cylindrical axis was taken along the $x$ axis.

For Fig.~\ref{figconfinedchi}(a) we initialized the director field as
\begin{eqnarray}
n_x &=& 0\\ \nonumber
n_y &=& \cos(-\phi/2+q_0x)\\ \nonumber
n_z &=& \sin(-\phi/2+q_0x),
\end{eqnarray}
while for Fig.~\ref{figconfinedchi}(b) 
\begin{eqnarray}
n_x &=& 0\\ \nonumber
n_y &=& \cos(\phi/2+q_0x)\\ \nonumber
n_z &=& \sin(\phi/2+q_0x).
\end{eqnarray}

Finally, for Fig.~\ref{figrotatingtau} we set, 
\begin{eqnarray}
n_x &=& \sin(q_0{\mathbf r}\cdot{\mathbf{\hat{h}}})\\ \nonumber
n_y &=&\cos(-\phi/2+q_0x)\cos(q_0{\mathbf r}\cdot{\mathbf{\hat{h}}})
\\ \nonumber
n_z &=&\sin(-\phi/2+q_0x)\cos(q_0{\mathbf r}\cdot{\mathbf{\hat{h}}}),
\end{eqnarray}
where ${\mathbf{r}}=(x,y,z)$ and ${\mathbf{\hat{h}}}=(\hat{h}_x,\hat{h}_y,\hat{h}_z)$, with
\begin{eqnarray}
\hat{h}_x &=& 0\\ \nonumber
\hat{h}_y &=& -\sin(\phi/2)\\ \nonumber
\hat{h}_z &=& \cos(\phi/2).
\end{eqnarray}

\subsubsection*{3D cholesterics}

For Fig.~\ref{figskyrmions}, we performed simulations for 3D bulk cholesterics, with periodic boundary conditions.  Parameters were set as follows: $\chi=3$ $q_0=\pi/16$, $A_0=0.005$, $L=0.01$, $\xi=0.7$, $\gamma=0.33775$ and the viscosity $\eta=5/3$. The simulation box was a parallelepiped with sizes $L_x=L_y=32$, $L_z=16$ in Fig.~\ref{figskyrmions}(a), and a cube with sizes $L_x=Ly=L_z=32$ in Fig.~\ref{figskyrmions}(b). \\

\subsection*{Simulated optical microscopy images}

The polarized optical microscopy images presented in Figs.~\ref{figconfinedchi} and~\ref{figrotatingtau} have been obtained following the procedure presented in~\cite{chen2024}, using the 
accompanying \textit{python} package \textit{LCPOM}, deposited in \url{https://github.com/depablogroup/lc-pom}.    

To generate single-wavelength (grayscale) micrographs from our simulation data, we begin by obtaining the tensor order parameter for each point in a regular grid. From the $\mathbf{Q}$
tensor, the pipeline extracts the local director and scalar order parameter $S$. It then computes the ordinary  refractive indices at the chosen wavelength ($500$ nm for all the cases considered, as no appreciable variation in the resulting pattern was observed at other wavelengths). Next,  the sample is discretised along the optical axis into a series of thin layers, each containing uniform orientation and birefringence. To capture how linearly polarized light propagates through these layers, the Jones matrix method is used~\cite{Ellis2019SimulatingOP}, multiplying the Jones matrices of all layers in succession. This approach determines how the phase and polarization state are changed within each layer, ultimately yielding, for every pixel, a final transmitted intensity measured between ideal crossed polarizers. The resulting two-dimensional intensity distribution thus corresponds directly to a grayscale polarized optical micrograph, facilitating comparison with experiments.

\if{
\appendix

\subsection*{Appendix: definitions of groups, rings, and algebras}

In this Appendix we give the definitions of groups, rings, and algebras, which are useful to some of the discussion in the main text.

First, a {\it group} is a set $G$ with one operation from $G\times G$ to $G$ (for instance addition or multiplication) which obeys some key specific properties -- associativity, existence of identity and existence of inverse. A {\it ring} is a set equipped with two operations, usually addition (which is commutative in the case of a ring) and multiplication. A ring is a group under addition and satisfies some of the properties of a group for multiplication -- namely associativity, and the existence of identity. Note a ring is not required to have a multiplicative inverse; a ring with a multiplicative inverse is called a {\it field}. An algebra is a vector space which has the properties of a ring. More formal definitions are given below.

{\bf Definition 1.} A {\it group} is a set G which is closed under an operation $*$ (i.e., for $x$ and $y$ $\in$ G, $x*y$ $\in$ G). For G to be a group, the following properties have to be satisfied:
\begin{itemize}
    \item {Identity. There exists $e$ $\in$ G, such that, for each $x$ $\in$ G, $x*e=e*x=x$.}
    \item{Inverse. For each $x$ $\in$ G, there exists $y$ such that $x*y=y*x=e$.}
    \item{Associativity. For each $x$, $y$, $z$ $\in$ G, $(x*y)*z=x*(y*z)$.
    }
\end{itemize}

{\bf Definition 2.} A {\it ring} is a set R, with two operations: addition (+) and multiplication ($\cdot$). For R to be a ring, three conditions have to be met. First, R has to be an abelian group for addition. Equivalently, the following properties have to be satisfied for addition:
\begin{itemize}
    \item{Additive associativity. For each $a$, $b$, $c$ $\in$ R, $(a+b)+c=a+(b+c)$.}
    \item{Commutative property. For each $a$, $b$ $\in$ R, $a+b=b+a$.}
    \item{Additive identity. There is an element $0$ $\in$ R such that, for each $a$ in R, $a+0=a$.}
    \item{Additive inverse. For each $a$ in R, there exists an additive inverse $-a$ also in R, such that $a+(-a)=0$.}
\end{itemize}
Second, 
the following properties have to be satisfied for multiplication:
\begin{itemize}
    \item{Multiplicative associativity. For each $a$, $b$ and $c$ $\in$ R, $(a\cdot b)\cdot c=a\cdot(b\cdot c)$.}
    \item{Multiplicative identity. There is an element $1$ in R such that, for each $a$ $\in$ R, $a \cdot 1=1 \cdot a=a$.}
\end{itemize}
Third, multiplication is distributive with respect to addition, or equivalently:
\begin{itemize}
    \item{Left distributivity. For each $a$, $b$ and $c$ $\in$ R, $a\cdot (b+c)=a\cdot c+b\cdot c$.}
    \item{Right distributivity. For each $a$, $b$ and $c$ $\in$ R, $(a+b)\cdot c=a\cdot c+b\cdot c$.}
\end{itemize}

{\bf Definition 3.} If $R$ is a ring, an algebra $A$ over $R$ is a vector space, where the binary operation (multiplication, denoted by $\cdot$) from $A \times A$ to $A$ satisfies compatibility with scalars. In formulas, this means that, for all elements $x, y\in A$, and for all $a, b\in R$,  $(ax)\cdot(by)=(ab)(x\cdot y)$.
Note sometimes $A$, as defined here, is referred to as an associative algebra (as multiplication is associative due to the definition of ring); in this case, multiplication is not assumed to be necessarily associative for a vector space to be called an algebra. Also note that multiplication is in general non-commutative. An example of a non-commutative algebra is that of 3D vectors with the cross product. Clifford algebras are another example, with the bilinear operator being the Clifford product -- e.g., sending $(e_i,e_j)$ into $e_i e_j$ (which can be viewed as matrix multiplication in the representations we have considered).

}\fi


%

\end{document}